\documentclass[aip,amsmath,amssymb,reprint,jap]{revtex4-1}

\usepackage{graphicx}
\usepackage{caption}
\usepackage{subcaption}
\usepackage{dcolumn}
\usepackage{bm}
\usepackage[utf8]{inputenc}
\usepackage[T1]{fontenc}
\usepackage[ngerman,main=english]{babel}
\usepackage{csquotes}
\usepackage{mathptmx}
\usepackage{etoolbox}
\usepackage{amsfonts,array}
\usepackage{siunitx} 
\usepackage{hyperref}
\usepackage{url}

\makeatletter

\def\@emailstar{%
  \@textsuperscript{\normalfont *}%
}

\def\@email#1#2{%
  \endgroup
  \expandafter\@mark@email\@author
  \patchcmd{\titleblock@produce}
    {\frontmatter@RRAPformat}
    {\frontmatter@RRAPformat{%
      \produce@RRAP{*#1\href{mailto:#2}{#2}}%
    }\frontmatter@RRAPformat}
    {}{}%
}

\def\@mark@email#1#2{%
  \def\@author{{#1\aftergroup\@emailstar}{#2}}%
}

\def\doauthor#1#2#3{%
  \begingroup
    \ignorespaces#1\unskip\@listcomma
    #3%
    \@if@empty{#2}{%
      \endgroup{}{}%
    }{%
      \endgroup{\comma@space}{}\frontmatter@footnote{#2}%
    }%
  \space\@listand
}

\makeatother

\begin{document}

\preprint{AIP/123-QED}


\title[Machine Learning Approaches for Improved Scalability of Metallic Magnetic Calorimeters]{Machine Learning Approaches for Improved Scalability of Metallic Magnetic Calorimeters}

\newif\ifarxivversion
\arxivversiontrue   
\newif\ifisacceptd
\isacceptdtrue      

\newif\ifisfirst
\isfirstfalse       

\ifarxivversion
    \ifisacceptd
        \thanks{This article may be downloaded for personal use only. Any other use requires prior permission of the author and AIP Publishing. This article appeared in M. O. Herdrich \textit{et al.}, J. Appl. Phys. \textbf{140} (17) 2026, pp. 174501 and may be found at \url{https://doi.org/10.1063/5.0348962}. Copyright 2026 M.O. Herdrich, T. Mattis, G. Weber, D.A. Schnauß-Müller, J.H. Walch, and Th. Stöhlker. This article is distributed under a Creative Commons Attribution (CC BY) License.}%
    \else
        \ifisfirst
            \thanks{Copyright 2026 M.O. Herdrich, T. Mattis, G. Weber, D.A. Schnauß-Müller, J.H. Walch, and Th. Stöhlker. This article is distributed under a Creative Commons Attribution (CC BY) License.}%
        \else
            \thanks{The following article has been submitted to Journal of Applied Physics. Copyright 2026 M.O. Herdrich, T. Mattis, G. Weber, D.A. Schnauß-Müller, J.H. Walch, and Th. Stöhlker. This article is distributed under a Creative Commons Attribution (CC BY) License.}%
        \fi
    \fi
\fi

\author{M.O. Herdrich}
    \email{m.o.herdrich@hi-jena.gsi.de}
    \affiliation{GSI Helmholtz Centre for Heavy Ion Research, Planckstr. 1, 64291 Darmstadt, Germany}
    \affiliation{Helmholtz-Institute Jena, Fröbelstieg 3, 07743 Jena, Germany}
    \affiliation{Institute of Optics and Quantum Electronics, Friedrich-Schiller-University Jena, Max-Wien-Platz 1, 07743 Jena, Germany}
   
\author{T. Mattis}
    \affiliation{Hasso-Plattner-Institute, Prof.-Dr.-Helmert-Straße 2-3, 14482 Potsdam-Babelsberg, Germany}

\author{G. Weber}
    \affiliation{GSI Helmholtz Centre for Heavy Ion Research, Planckstr. 1, 64291 Darmstadt, Germany}
    \affiliation{Helmholtz-Institute Jena, Fröbelstieg 3, 07743 Jena, Germany}

\author{D.A. Schnauß-Müller}
    \affiliation{GSI Helmholtz Centre for Heavy Ion Research, Planckstr. 1, 64291 Darmstadt, Germany}
    \affiliation{Helmholtz-Institute Jena, Fröbelstieg 3, 07743 Jena, Germany}
    \affiliation{Institute of Optics and Quantum Electronics, Friedrich-Schiller-University Jena, Max-Wien-Platz 1, 07743 Jena, Germany}

\author{J.H. Walch}
    \affiliation{GSI Helmholtz Centre for Heavy Ion Research, Planckstr. 1, 64291 Darmstadt, Germany}
    \affiliation{Helmholtz-Institute Jena, Fröbelstieg 3, 07743 Jena, Germany}
    \affiliation{Institute of Optics and Quantum Electronics, Friedrich-Schiller-University Jena, Max-Wien-Platz 1, 07743 Jena, Germany}

\author{Th. Stöhlker}
    \affiliation{GSI Helmholtz Centre for Heavy Ion Research, Planckstr. 1, 64291 Darmstadt, Germany}
    \affiliation{Helmholtz-Institute Jena, Fröbelstieg 3, 07743 Jena, Germany}
    \affiliation{Institute of Optics and Quantum Electronics, Friedrich-Schiller-University Jena, Max-Wien-Platz 1, 07743 Jena, Germany}

\date{\today}

\begin{abstract}
    \textit{MMCs} (Metallic Magnetic Calorimeters) are a promising tool for high-precision X-ray spectroscopy. However the signal processing and calibration pipeline required to exploit their full potential is complex and difficult to scale for large, highly pixelated detector arrays. In this work, we investigate \textit{ML} (Machine Learning) methods for \textit{maXs}-series detectors, focusing on pulse classification and artifact rejection as well as pulse-shape analysis and feature extraction. By combining unsupervised label auto-discovery with supervised classification and regression, we obtain robust and scalable artifact rejection for high-confidence template building. For ML-based feature extraction, our results constitute a promising proof of principle: in simulation, the ML estimator remains competitive across realistic pulse-shape and noise variation. On real detector data, the method reaches performance comparable to the hand-optimized FIR baseline before gain-drift correction, while further domain adaptation is required to match the established manual analysis pipeline. Both results represent important steps toward a fully automated and scalable data processing procedure.
\end{abstract}

\maketitle


\section{Introduction}

Benchmarking our understanding of elementary physical processes, such as bound-state quantum electrodynamics in atomic and fundamental physics, places highest demands on the precision that can be achieved in connected experiments. In recent years, micro-calorimeter detectors based on \textit{MMC} (Metallic Magnetic Calorimeters) technology have developed into promising tools for high-precision spectroscopy: They are employed, e.g., in neutrino-mass experiments (ECHo \cite{Mantegazzini2022ECHo1k}), precision atomic physics \cite{piesMaXsMicrocalorimeterArrays2012,Krantz_2024_APL}, laboratory astrophysics (Lynx \cite{Lynx_XMS_PMC}), and ultra-fast X-ray spectroscopy of molecular systems \cite{Uhlig2013PRL}.

Micro-calorimeters work by absorbing the energy of an incident single photon giving rise to a proportional change in sensor temperature \cite{fleischmannMetallicMagneticCalorimeters2005}. This is measurable via a temperature-dependent material property, e.g., in the case of MMCs the changing magnetization of a paramagnetic \textit{Ag:Er} (dilute alloy of erbium in silver) sensor in conjunction with highly sensitive current-sensing \textit{SQUIDs} (Superconducting Quantum-Interference Devices) \cite{kempfDirectcurrentSuperconductingQuantum2015}. Due to their unique working principles MMCs combine several advantageous properties of conventional energy- and wavelength-dispersive X-ray photon detectors: Similar to semiconductor detectors, they are capable of covering a broad spectral bandwidth with a high quantum-efficiency~\textendash~tunable via the thickness and material choice of a dedicated absorber~\textendash~typically in the range of \SI{1}{\kilo\electronvolt} to above \SI{100}{\kilo\electronvolt}. At the same time they reach an excellent energy resolution that~\textendash~depending on their application~\textendash~is at least an order of magnitude better than that of semiconductor detectors. Notable examples are \SI{10}{\electronvolt} \textit{FWHM} (full width at half maximum) at \SI{60}{\kilo\electronvolt} photon energy \cite{geistBestimmungIsomerenergie229Th2020} or \SI{1.25}{\electronvolt} at \SI{5.9}{\kilo\electronvolt} \cite{PhysRevD.109.043035} across their respective spectral range. Furthermore, MMCs exhibit excellent spectral linearity with non-linearities well understood form first principles \cite{piesMaXsMicrocalorimeterArrays2012}. Due to their fast rise time \cite{fleischmannMetallicMagneticCalorimeters2005}, they can also be used in coincidence measurement schemes in conjunction with other detectors \cite{pfaffleinExploitationTimingCapabilities2022}. In this work, we focus on the \textit{maXs} (MMC Array for X-ray Spectroscopy)-series detectors developed within the \textit{SPARC}-collaboration \cite{SPARC_maXs_TDR_2016}. They have already been employed successfully in several benchmark experiments \cite{herdrichXraySpectroscopyBased2023, herdrichHighresolutionXrayEmission2023, herdrichApplicationMetallicmagneticCalorimeter2024} and dedicated high precision measurements on the campus of GSI/FAIR \cite{pfaffleinQuantumElectrodynamicsStrong2025}, demonstrating their excellent performance in the challenging environment of heavy-ion storage rings.
\begin{figure}[ht!]
    \centering
    \includegraphics[width=0.45\textwidth]{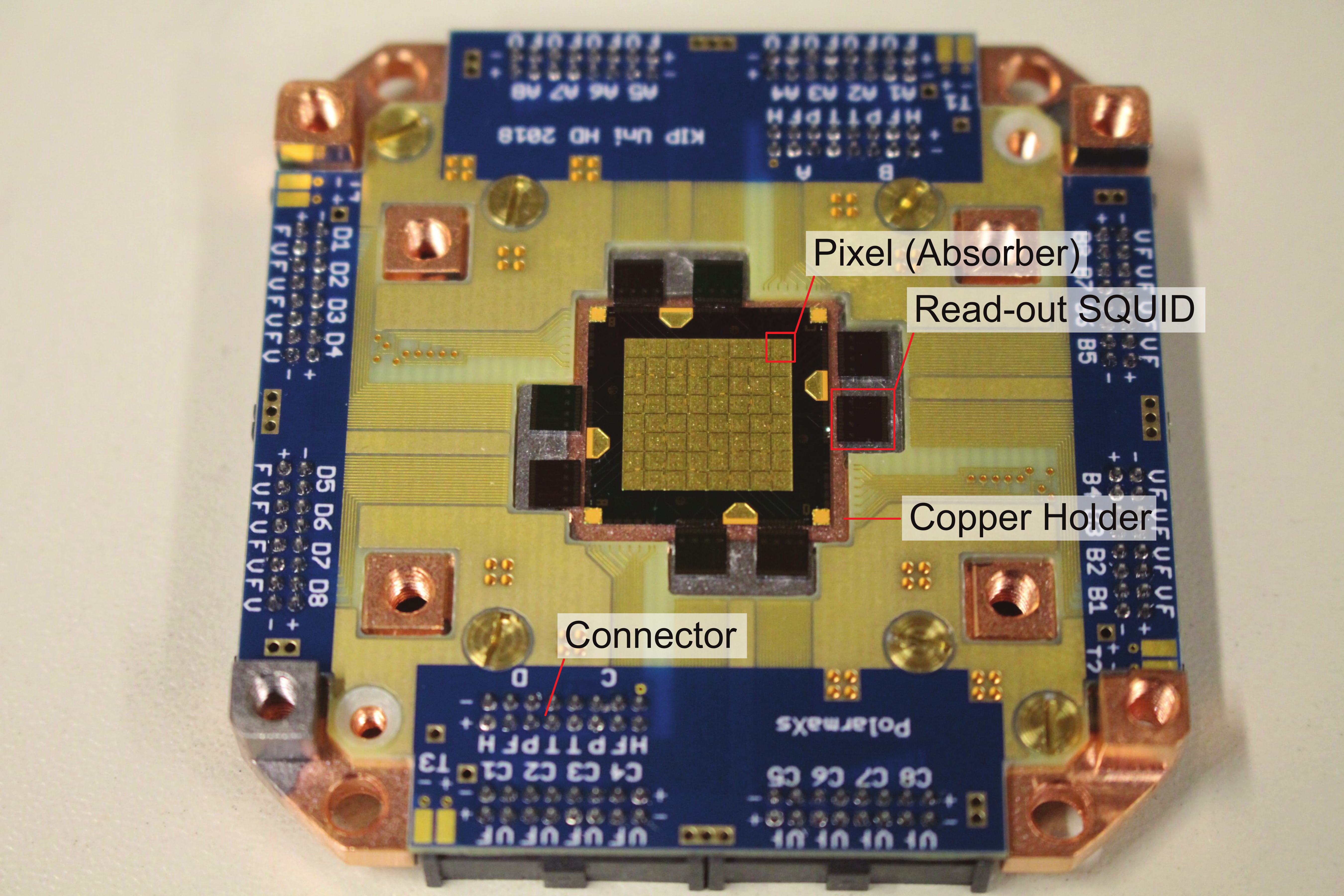}
    \caption{Photograph of a maXs-100 detector chip featuring 64 pixels. The absorbers of the individual detector pixels can be seen as small golden squares in the center of the chip. The attached sensors are located underneath the absorbers. The chip is mounted directly to the copper holder and connected to the read-out electronics via wire bonds (not yet attached in the picture).}
    \label{fig:MaXs-100 Detector Chip}
\end{figure}

However, utilizing these detectors requires overcoming several challenges to exploit their full potential. At typical heat capacities of soft- to hard X-ray detectors of around \SIrange[range-units=single]{1}{10}{\pico\joule\per\kelvin} \cite{hengstlerDevelopmentCharacterizationTwodimensional2017} the expected temperature change in the sensor only amounts to an order of magnitude of \SIrange[range-units=single]{100}{10}{\micro\kelvin\per\kilo\electronvolt}. The material properties required for the sensor to combine low heat capacity with sufficient thermal conductivity necessitate operation at temperatures below approximately 20 mK. Dedicated hardware~\textendash~like a dilution cryostat to reach temperatures below \SI{20}{\milli\kelvin}, the SQUID read-out and amplification stages, etc.~\textendash~is required for their operation. Furthermore, the high sensitivity of the sensor leads to an increased susceptibility to external sources of noise, e.g., acoustic vibrations or external electromagnetic fields. Resulting artifacts, like the temperature sensitive drift of the sensor's gain (see, for example, M.O. Herdrich et al.\cite{herdrichXraySpectroscopyBased2023}), necessitate the shift from traditional analog signal processing to a digital pulse-shape analysis chain to mitigate these effects. Therefore, for maXs detectors, extensive measurement and calibration procedures and software packages have been developed \cite{hengstlerDevelopmentCharacterizationTwodimensional2017,herdrichAnwendungKryogenerKalorimeter2023} dealing with pulse classification, pulse-shape analysis and artifact correction. These methods include, for example, continuous calibration schemes \cite{pfaffleinQuantumElectrodynamicsStrong2025} and \textit{FIR} (Finite Impulse Response) filters specialized for different use-cases, like the \textit{MWD} (Moving Window Deconvolution) for highest stability in noisy experiment environments \cite{herdrichAnwendungKryogenerKalorimeter2023}, or the optimal filter designed to yield the highest energy resolution \cite{fleischmannMetallicMagneticCalorimeters2005}.

The individual components of the detector operation and read-out chain require careful optimization of the involved  parameters in both soft- and hardware to achieve optimal performance. This comprises, for example, the tuning of the SQUID amplification stages, or the adaptation of the pulse-shape analysis to the specific sensor characteristics. So far, these procedures could only be partially automated and are, as a consequence, labor- and time-intensive. Furthermore, to achieve highest signal-to-noise ratios, MMC detectors are designed to have the lowest heat capacities possible. Since material choices are constrained by competing requirements~\textendash~like good thermal conductivity~\textendash~and the thickness of the absorber is determined by the desired quantum efficiency for a given photon energy, shrinking the absorber area is usually the only available parameter to minimize the heat capacity. To still guarantee an acceptable active area of the detector, MMC detector systems, such as the \textit{maXs}-series detector family, are pixelated and consist of several absorber-sensor pairs acting as individual detectors on a single chip (see, for example, fig.~\ref{fig:MaXs-100 Detector Chip}). Slight variations of the sensor characteristics make a separate optimization step for each detector channel necessary. For $64$ pixels of the currently used maXs-30 and maXs-100 detectors this is already challenging. However, in the future detector systems with thousands of pixels are planned (see, for example, M. Wegner et al.\cite{wegnerMicrowaveSQUIDMultiplexing2018}), rendering the current manual setup and optimization chain virtually impossible to apply. Additionally, the traditional processing pipelines rely on fixed heuristics that might not be sufficient for handling every edge case that could result from this complexity. Thus, new methods are required to make the process more efficient and scalable for future applications.

In this paper, we will first discuss what \textit{ML} (Machine Learning) means in the context of MMCs and how it can be applied to address the challenges mentioned above. We will then present two example applications of ML methods within the context of our pulse processing pipeline: a pulse classifier for artifact rejection and a pulse-shape analysis method for feature extraction. Finally, we will conclude with a discussion of the results and an outlook on future work in this area.


\section{Machine Learning in the Context of MMCs}
\begin{figure}[ht!]
    \centering
    \includegraphics[width=0.45\textwidth]{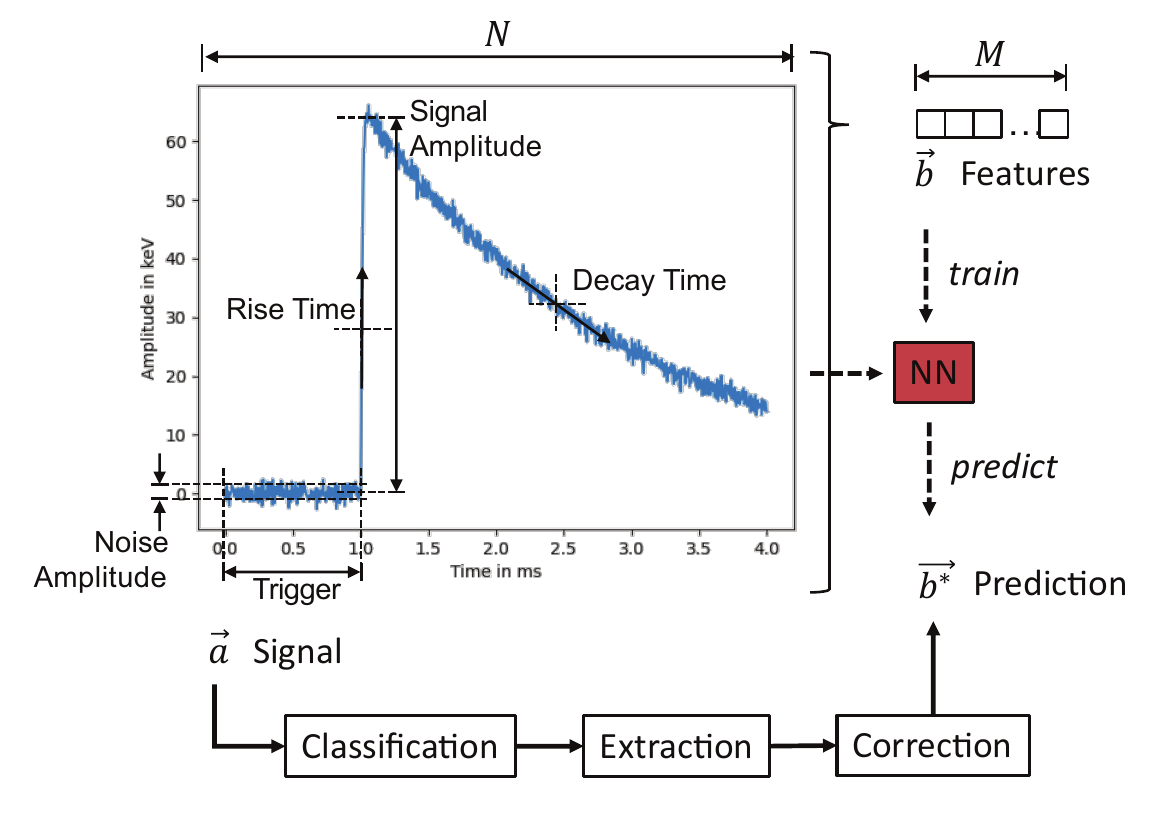}
    \caption{Abstract representation of the signal processing pipeline used to extract all relevant signal features from the MMC pulses. In the top left the trace of a typical MMC pulse, i.e., the digitized time-domain voltage response to a photon absorption event, is shown. Each raw waveform represents an $N$-dimensional vector $\vec{a}$. The most relevant features of the pulse are highlighted in the plot. The reduced set of physical parameters forms an $M$-dimensional vector $\vec{b}$, with $M \ll N$. The goal of the processing pipeline (bottom) is to find a mapping $\vec{a} \mapsto \vec{b}$ that extracts all relevant features from the raw pulse. The pipeline consists of several steps, including pulse classification, feature extraction, and corrections (like gain-drift corrections and calibration~\textendash~requiring additional data). For each step, ML tools (for example a NN) can be trained to learn aspects of this mapping, by using sets of labeled signals (right). These ML tools could complement the traditional processing methods to increase the robustness and scalability of the pipeline.}
    \label{fig:ML Pipeline MMC}
\end{figure}
In contrast to the techniques discussed so far, ML is a mostly data-driven modeling approach rather than rule-based programming \cite{bishopPatternRecognitionMachine2006}. In the context of detector signal analysis, it discovers functional relationships that map raw detector signals to the physically meaningful quantities they encode. Particularly, deep learning enables the automatic extraction of hierarchical representations that capture increasingly abstract physical features \cite{10.5555/303568.303704}. \textit{NN} (Neural Networks) thereby act as flexible function approximators that non-linearly map the high-dimensional space of measured data to a set of relevant features. For the system discussed in this paper, this amounts to $2^{14}$ samples to single-digit counts of meaningful parameters per recorded signal. The process requires a training procedure that adapts the NN to the specific problem at hand. Supervised learning methods utilize the existence of known labels to translate a set of given inputs, for example, to estimated pulse parameters. Contrary to that, unsupervised learning describes ways of discovering structures without predefined labels \cite{jingSelfsupervisedVisualFeature2019}.

While ML offers a broad range of promising methods and has become increasingly prominent across many scientific disciplines, its application must be guided by a clear physical motivation. In the present context, conventional analysis techniques face intrinsic limitations arising from the lack of labeled data, the complexity of the detector response and its environmental dependence, and the need for scalable processing. In recent years, ML approaches have proven highly successful for similar applications across other physical sciences \cite{carleoMachineLearningPhysical2019}. The data-driven nature of ML enables automatic pattern discovery in large datasets and the interpretation of complex systems. Particularly for highly pixelated detector arrays these methods could prove advantageous as they are less reliant on hand-tuned parameters. ML architectures can learn corrections to imperfect physics models and capture non-linear effects and correlations that are difficult to model or describe analytically otherwise. Furthermore, because of the highly optimized implementation of most ready-made ML frameworks (see, for example, \textit{PyTorch} \cite{paszkePyTorchImperativeStyle2019} or \textit{scikit-learn} \cite{scikit-learn} for the \textit{python} programming environment), once trained, the inference is extremely fast. This makes ML suitable for large-scale detector arrays even in online monitoring scenarios. Finally, ML methods ideally are meant to complement~\textendash~not replace~\textendash~the traditional calibration and filtering pipelines (see fig.~\ref{fig:ML Pipeline MMC}).

Beyond the question of applicability, several other challenges have to be overcome in the utilization of ML for MMCs. The measured data shows a high heterogeneity due to pixel-to-pixel variations, and drifts or sudden jumps in the measurement parameters over time. Therefore, the utilized architectures need to be robust and have an intrinsic tolerance to these fluctuations. Intermediate states and features of ML methods are often abstract and need targeted validation and careful design to be interpretable in physically meaningful terms \cite{liptonMythosModelInterpretability2017}. Additionally, NNs tend to run into overfitting or exploit spurious correlations, where they might learn setup-specific artifacts instead of physics \cite{zhangUnderstandingDeepLearning2017}. To mitigate this, careful regularization and validation are needed. Most importantly, though, the lack of a labeled ground truth poses the largest challenge. No "true" values are available for physical quantities in measured data. Yet we want models that predict these parameters for us, leading to an apparent paradox. Simulation models might not perfectly match real detectors and manually labeling data sets tends to be impractical or not feasible without significant errors. This renders the direct application of supervised learning schemes impossible. Particularly for hardware optimization, feedback loops can be very time consuming or difficult to correlate, which is why this work focuses on software-related issues first.

We have devised several strategies to tackle these challenges: Starting from raw detector pulses as inputs it is possible to learn representations of the data via self-supervised training like auto-encoders or to derive features via auto-discovery methods like clustering. This enables us to move to supervised learning models in subsequent stages of the pipeline or to use transfer learning methodology in conjunction with simulated data to bridge the gap \cite{tobinDomainRandomizationTransferring2017, panSurveyTransferLearning2010}. Finally, validation via simulations with a wide range of parameters helps to cover potential edge cases and allows for more physical interpretability when applied to real data later. The following section demonstrates our approach with two example applications of ML methods to the MMC processing pipeline.


\section{Application of Machine Learning Methods}


\subsection{Example 1: Pulse Classifier and Artifact Rejection}
\begin{figure} [ht!]
    \centering
    \includegraphics[width=0.35\textwidth]{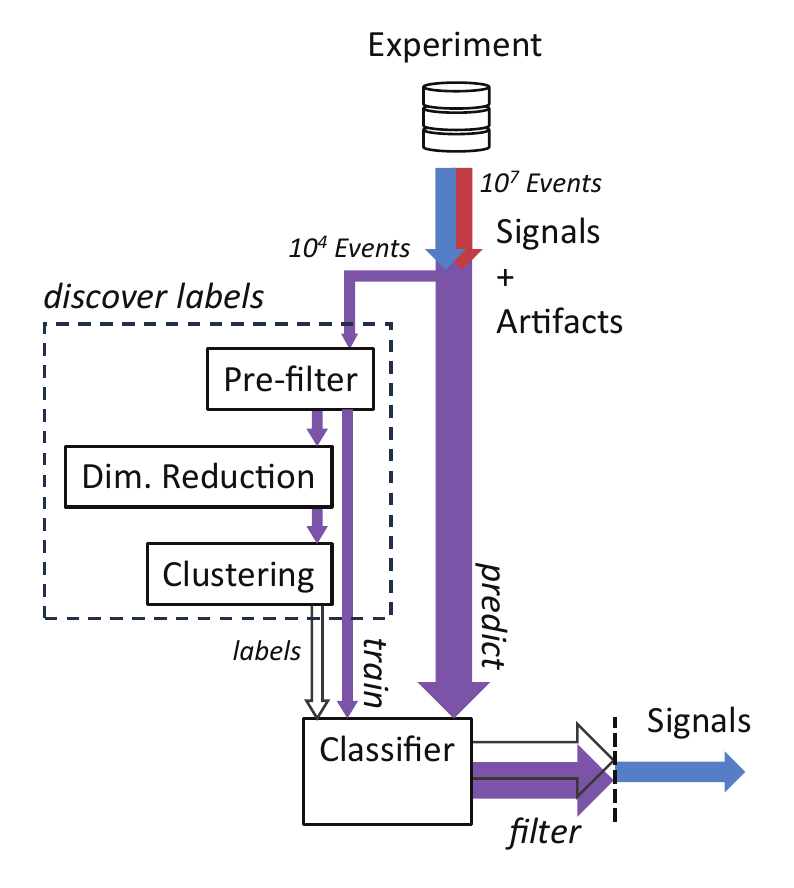}
    \caption{Schematic representation of the Reduce-Cluster-Classify pipeline for label auto-discovery in pulse classification. Traces corresponding to valid pulses (blue arrows) and artifacts (red arrows) are loaded from the recorded experiment data and the mixed population (purple arrows) is processed by the analysis pipeline. A portion of traces (around $1000$ events per ADC channel) is split off and used for the auto-discovery of classification labels via dimensionality reduction and clustering. The subset together with the corresponding labels (white arrows) can then be utilized, for example, to train a supervised classifier. Finally, the resulting classifier can be applied to the full dataset. This way artifacts can be detected and filtered out before the next analysis step.}
    \label{fig:Reduce-Cluster-Classify Pipeline}
\end{figure}
Due to the high sensitivity of MMCs, their performance might be affected by noise issues caused by external sources. In particular, the recorded traces occasionally contain artifacts that trigger the data acquisition and produce a signal, but do not correspond to events that we intend to measure. In general, artifacts are defined as any pulse-shape that deviates from the expected detector response upon absorption of an X-ray photon (see, for example, the pulse-shape and artifact depicted in fig.~\ref{fig:UMAP HDBSCAN Example}). They can have a multitude of sources and range from arbitrary signals~\textendash~likely due to the capture of electromagnetic noise like wireless network carriers~\textendash~to distorted pulses, i.e., overshoots of the peak, significantly altered rise and decay times, etc. The latter might originate, for example, from events in which the sensor below the absorber was directly hit by a photon or from charged particles hitting the absorber. The correspondence between a specific artifact and its origin is not necessarily always known. The presence of these artifacts, however, can severely degrade the measurement performance of the detector~\textendash~if they are not properly identified and filtered out~\textendash~as they cannot be properly processed by the subsequent pulse-shape analysis. One of the first tasks of the pulse processing pipeline is therefore to reliably separate valid pulses from artifacts. Traditional methods rely on simple thresholding and shape-based cuts that are not able to capture the full complexity of the data. Furthermore, these methods require extensive manual tuning and are not easily scalable to large pixel counts. The occurrence of artifacts is particularly problematic for the generation of template pulses, i.e., representative detector responses reconstructed from ensembles of valid events, which are, for example, used in the construction of optimal filters. While pulse-shape variations between valid events within each detector channel are expected to be marginal, artifacts can significantly bias the resulting template. Threshold-based methods can be used to filter out the most obvious artifacts, however, more complex and subtle cases require a more sophisticated approach.

One of the most common applications of ML is the classification of instances, like images or pulses in our case, and many different classifier architectures are available. However, most of these methods require labeled training data which is not available in our case. Manually labeling the data is impractical and error-prone due to the large data volumes and the complexity of the artifacts. Instead, we can exploit the fact that valid pulses and artifacts follow recognizable patterns that can be discovered via unsupervised learning methods. This allows us to generate our own labels in a self-supervised manner and, for example, subsequently train a supervised classifier on these labels (see fig.~\ref{fig:Reduce-Cluster-Classify Pipeline}). In the context of \textit{CV} (computer vision), a common approach to this problem is the Reduce-Cluster-Classify pipeline (see, for example, DeepCluster\cite{carleoMachineLearningPhysical2019}).


\subsubsection{Dimensionality Reduction}

Analogous to CV problems, our data is also highly dimensional and sparse, making it difficult to identify patterns directly. Therefore, a dimensionality reduction step is applied to project the data into a lower-dimensional space where structural patterns emerge more clearly. This process is implemented utilizing two commonly available algorithms. First, \textit{PCA} (Principal Component Analysis) \cite{pearsonLIIILinesPlanes1901} is used to reduce the data to a few hundred dimensions. PCA identifies the directions (axes) along which the data varies the most and projects it onto these axes, retaining as much information as possible with fewer dimensions. This method is particularly effective at capturing linear correlations in the data and extracting global features. Next, \textit{UMAP} (Uniform Manifold Approximation and Projection) \cite{mcinnesUMAPUniformManifold2020} is applied for a final reduction to two dimensions. UMAP learns a low-dimensional layout that preserves both local and global structure by assuming the data lies on a curved manifold. Its non-linear nature makes it well-suited for capturing local features in the data.


\subsubsection{Clustering}
\begin{figure}[ht!]
    \centering
    \includegraphics[width=0.45\textwidth]{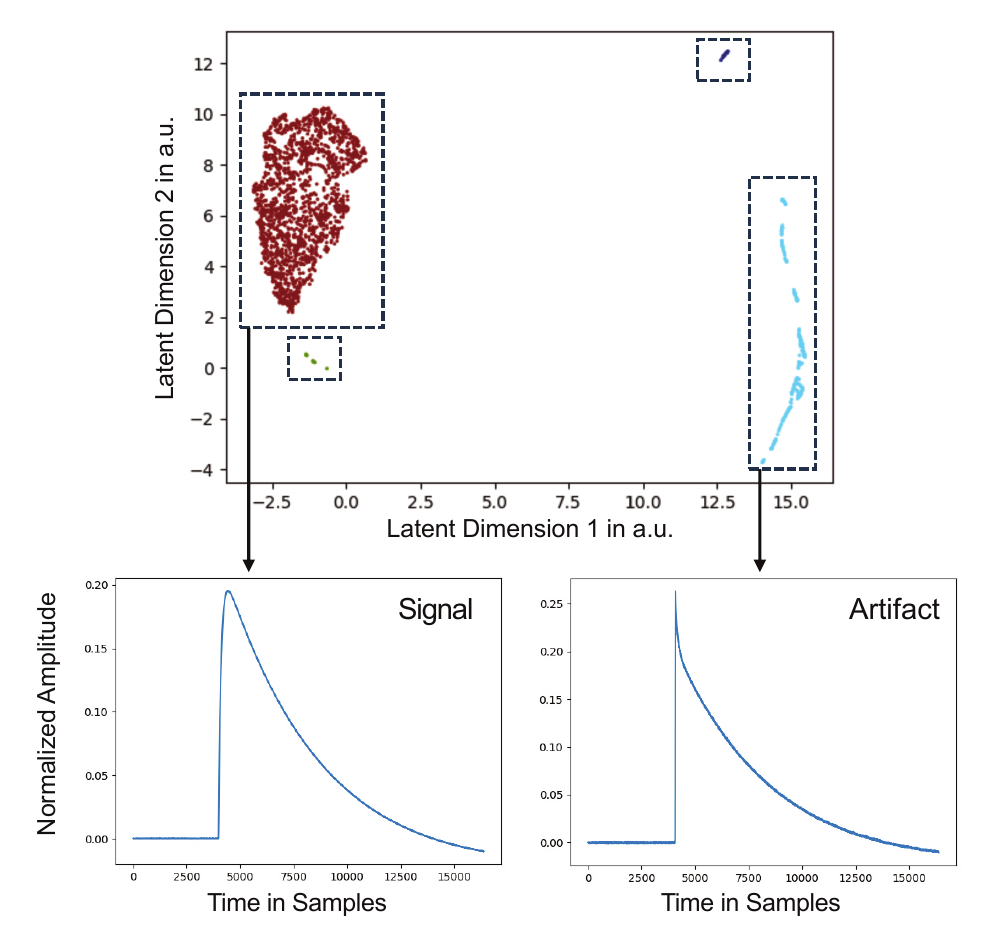}
    \caption{This example shows the combined PCA and UMAP projection of signal data with HDBSCAN clustering results in the plot at the top. Each point corresponds to one trace after embedding. The axes are latent embedding coordinates and have no direct physical interpretation. Each cluster identified by HDBSCAN is indicated by a different color. The largest cluster corresponds to valid pulses, while all other clusters represent artifacts. The assignment of the dominant cluster as the pulse population corresponding to the undistorted detection of an X-ray photon event is based on its waveform morphology and the assumption that artifacts are less common than valid events. The lower panel shows averaged traces from the dominant cluster and the largest artifact cluster. It can be seen that the artifact traces have a significantly different shape compared to the undistorted pulse, potentially resulting from a direct hit of the sensor by a photon.}
    \label{fig:UMAP HDBSCAN Example}
\end{figure}
The second step involves clustering the reduced data to identify groups of similar patterns. This is achieved using a hierarchical density-based clustering algorithm \textit{HDBSCAN} (Hierarchical Density Based Scan) \cite{campelloHierarchicalDensityEstimates2015}. Groups of data points that are closely packed together in dense regions are identified as clusters, while sparse regions are marked as noise. HDBSCAN was chosen over other clustering algorithms like \textit{DBSCAN} \cite{esterDensitybasedAlgorithmDiscovering1996} due to its reduced sensitivity to hyper-parameters and its ability to produce more stable clustering results. It also allows for soft clustering, where each data point is assigned a probability of belonging to each cluster rather than a hard assignment. An exclusion threshold can be assigned later to filter out uncertain assignments and thus further increase the confidence in the identified clusters. This is particularly important for the creation of a reliable template pulse, where the quality of the selected pulses is more important than their quantity.\begin{figure*}[t]
    \centering
    \includegraphics[width=0.82\textwidth]{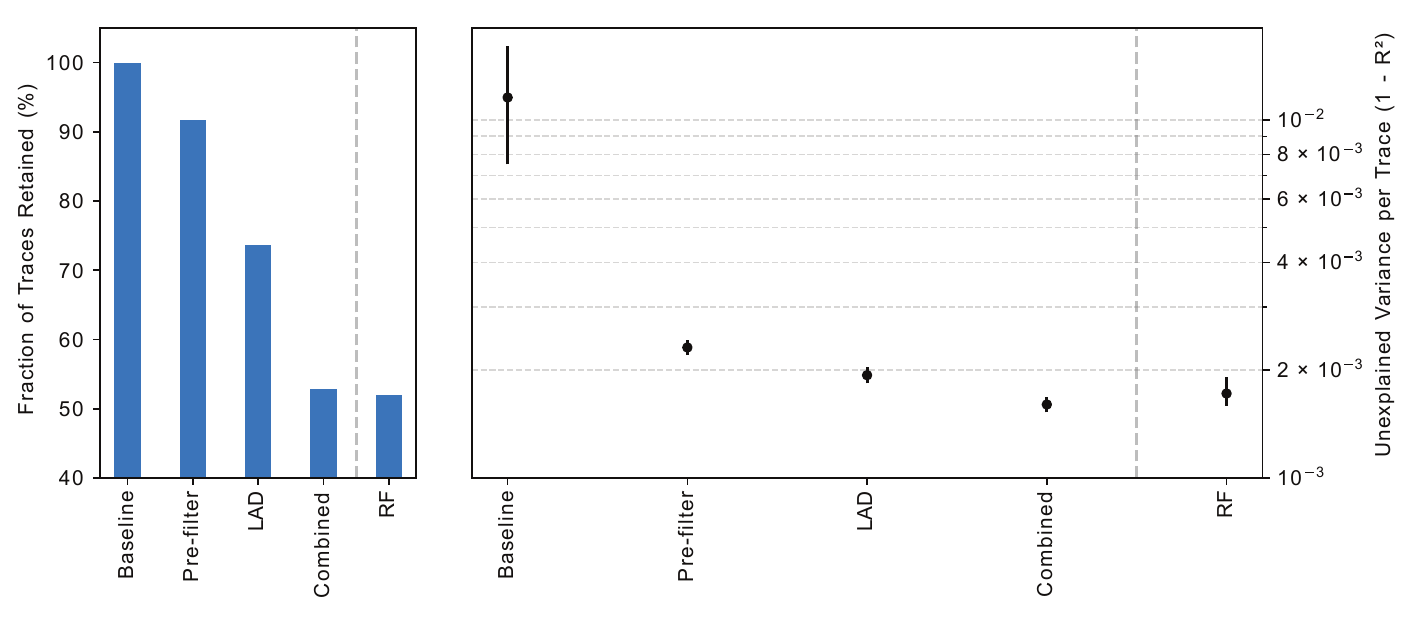}
    \caption{On the left side, the fraction of retained pulses $f$ is shown for the different selection methods applied to the same dataset. The method labeled "Baseline" refers to the unfiltered pulse population, "Pre-filter" to the traditional threshold-based selection, "LAD" to the label auto-discovery method, "Combined" to the combination of pre-filter and LAD, and "RFC" to a random forest classifier trained on the labels generated by the full pre-filter and LAD pipeline. On the right side, the global unexplained variance ($1-R^2$) between the selected pulses and the matched common template is shown. The error bars indicate the \SI{95}{\percent} confidence interval as determined by bootstrapping the selected pulse population. The LAD method retains a smaller fraction of pulses than the traditional pre-filter, but yields a lower unexplained variance, indicating a more consistent pulse population. The combination of both methods further improves the template quality.}
    \label{fig:LAD comparison}
\end{figure*}

Finally, the largest cluster is identified as the group of valid pulses, while all other clusters are considered artifacts (see fig.~\ref{fig:UMAP HDBSCAN Example}). This assumption is based on the expectation that valid events occur more frequently than artifacts in the data. By repeating the label auto-discovery process several times with different initial randomization seeds, multiple labels can be found for a single trace. A majority vote across all found labels can then be utilized to reduce the influence of outliers and increase the robustness of the labeling process. Labels found by this method can be merged with pre-filter results, like simple leading edge triggers and threshold cuts, to further improve the classification performance.

The PCA algorithm is available in the scikit-learn library, while UMAP and HDBSCAN are implemented using their respective open-source libraries. The performance of the label auto-discovery step can be evaluated qualitatively by visual inspection of randomly selected traces from each cluster. Additionally, the stability of the clustering results is assessed by measuring the consistency of labels across multiple runs with different random seeds. Evaluation metrics readily available in scikit-learn like the \textit{silhouette score} and the \textit{NMI} (Normalized Mutual Information) or \textit{ARI} (Adjusted Rand Index) can be used to quantify the quality of the clustering results for additional diagnostics. A high silhouette score indicates that the clusters are well-separated and compact, while high NMI or ARI values suggest that the clustering results are consistent across different runs. Overall, this auto-discovery approach allows for the generation of reliable labels for pulse classification without the need for manual labeling, enabling the subsequent creation of a classifier. In particular, the label auto-discovery approach allows us to identify artifacts that we do not need to predefine. While threshold filters only remove a limited number of expected deviations from the undistorted detector response, the label auto-discovery method can identify a wide range of artifacts based on their inherent patterns in the data. 

\begin{table*}[t]
    \centering
    \caption{Template quality metrics for the different pulse-selection methods. Values are given as mean with the corresponding \SI{95}{\percent} confidence interval in parentheses (as determined by bootstrapping the selected pulse population). The RMS template variation $V$ is given in normalized amplitude units (n.u.), i.e., the full \textit{ADC}'s (analog-to-digital converter) sample range normalized to values between $-1$ and $1$. The other quantities are dimensionless. The method labeled "Baseline" refers to the unfiltered pulse population, "Pre-filter" to the traditional threshold-based selection, "LAD" to the label auto-discovery method, "Combined" to the combination of pre-filter and LAD, and "RFC" to a random forest classifier trained on the labels generated by the full pre-filter and LAD pipeline. In the shown example a total population of $M_{\text{total}} = 1600$ pulses was used.}
    \label{tab:template_quality}
    \begin{tabular}{lcccc}
        \hline
        Method &
        Retained fraction ($f$) &
        Template variation ($V$) [$\times 10^{-4}$ n.u.] &
        Unexplained Variance ($1-R^2$) [$\times 10^{-3}$] &
        Residual / Noise ($S$) \\
        \hline
        Baseline &
        $1.00$ &
        $5.33\;(3.92\mathbin{-}9.21)$ &
        $11.5\;(7.51\mathbin{-}16.1)$ &
        $1.23\;(1.15\mathbin{-}1.33)$ \\
        Pre-filter &
        $0.917$ &
        $2.77\;(2.54\mathbin{-}3.21)$ &
        $2.31\;(2.20\mathbin{-}2.43)$ &
        $1.01\;(1.01\mathbin{-}1.02)$ \\
        LAD &
        $0.737$ &
        $2.76\;(2.51\mathbin{-}3.40)$ &
        $1.94\;(1.84\mathbin{-}2.04)$ &
        $1.01\;(1.01\mathbin{-}1.02)$ \\
        Combined &
        $0.529$ &
        $2.54\;(2.27\mathbin{-}3.20)$ &
        $1.60\;(1.52\mathbin{-}1.69)$ &
        $1.01\;(1.00\mathbin{-}1.01)$ \\
        \hline
        RFC &
        $0.520$ &
        $2.70\;(2.32\mathbin{-}3.37)$ &
        $1.72\;(1.59\mathbin{-}1.92)$ &
        $1.03\;(1.02\mathbin{-}1.04)$ \\
        \hline
    \end{tabular}
\end{table*}
\subsubsection{Classification}
The unsupervised reduce-cluster stage yields cluster labels that identify a high-confidence population of template-compatible detector responses and a complementary set of statistically inconsistent traces. In this work, we refer to this step as \textit{LAD} (Label Auto-Discovery). In practice, LAD labels can be used in two ways: (i) directly, by retaining only the high-confidence population for template construction, and (ii) indirectly, by training a supervised classifier that reproduces this selection at scale.

We compare the performance of the traditional pre-filter stage (consisting of a baseline-noise rejection, leading edge trigger, simple amplitude threshold cut, and a decay time cut) to the performance of the LAD-based approach. To quantify the performance of the different methods, we define several metrics: (i) We report on the fraction $f$ of the size of the pulse population retained by the respective selection procedure $M$ to the total number of pulses $M_{\text{total}}$, quantifying its selectivity:
$$
f = \frac{M}{M_{\text{total}}}
$$
(ii) By resampling multiple subsets of selected pulses (at \SI{80}{\percent} of the total selected population in our example), we can bootstrap a distribution of resulting template pulses and evaluate their waveform-to-waveform variance within the population. A template is built from the subset by aligning each pulse temporally, normalizing their amplitudes and calculating their point-wise average pulse-shape. The point-wise \textit{RMS} (Root-Mean-Square) variation of templates $V$ serves as a measure to quantify the statistical stability of the reconstructed detector response:
$$
V = \sqrt{\frac{1}{NM}\sum_{i=1}^{M}\sum_{j=1}^{N} \left( T_{ij} - \bar{T}_{j} \right)^2}
$$
Here $N$ is the total number of samples, $T_{ij}$ is the $j$-th sample of the $i$-th template and $\bar{T}_{j}$ is the mean of all templates at that sample. (iii) Additionally, we can evaluate the residuals of each selected pulse after subtracting the matched template. The global explained variance ($R^2$) is the fraction of the total variance of the selected pulse population explained by a common fitted template. It measures the internal consistency of the selected pulse-shapes:
$$
R^2 = 1 - \frac{\sum_{i=1}^{M} \sum_{j=1}^{N} \left( x_{ij} - \hat{T}_{ij} \right)^2}{\sum_{i=1}^{M} \sum_{j=1}^{N} \left( x_{ij} - \bar{T}_{j} \right)^2}
$$ Here $x_{ij}$ is the $j$-th sample of the $i$-th selected pulse, $ \bar{T}_{j}$ is the corresponding sample of the common template (built from all pulses in the selected population), and $\hat{T}_{ij} = a_{i} \bar{T}_{j-j_i} + b_{i}$ is the common template matched to the $i$-th pulse (with amplitude $a_i$, baseline $b_i$, and pulse arrival time $j_i$). (iv) Finally, we can compare the RMS of the residuals after subtracting the matched templates to the RMS of the detector noise itself. This gives us as a measure of how closely the remaining pulse-to-pulse deviations approach the intrinsic noise floor:
$$
S = \sqrt{\frac{K\sum_{i=1}^{M} \sum_{j=1}^{N} \left( x_{ij} - \hat{T}_{ij} \right)^2}{M\sum_{i=1}^{K} \sum_{j=1}^{N} \left( n_{ij} \right)^2}}
$$ Where $n_{ij}$ is the $j$-th sample of the $i$-th of $K$ independently measured pulse-free noise traces. The distribution of both the explained variance and the noise ratio, can be estimated by bootstrapping multiple subsets of the selected pulse population as well.

In the example dataset presented in this work we selected $M_{\text{total}}\approx1600$ traces. A preliminary scan of the dataset size indicated that, beyond this point, further increases in the number of traces yielded only marginal improvements in the resulting template quality, while substantially increasing the computational cost. The results of the comparison between the traditional pre-filter and the LAD method are shown in fig.~\ref{fig:LAD comparison} and table~\ref{tab:template_quality}. The traditional filter removes known pathological events, but leaves a substantial population of subtle waveform variations. Compared to the baseline~\textendash~i.e., unfiltered data~\textendash~it achieves a reduction of the RMS template variation by almost half and the amount of unexplained variance is reduced by \SI{80}{\percent}. LAD discovers statistically inconsistent pulse populations without requiring those deviations to be specified a priori. It reaches a template variation on the same level of stability as the traditional filter, however, manages to find a slightly more statistically consistent population of selected pulses, reducing the unexplained variance by \SI{83}{\percent}. The sequential combination of the pre-filter and LAD yields the most stable and internally consistent template population (unexplained variance is reduced by \SI{86}{\percent}), with a residual structure between the individual pulses and the matched templates approaching the intrinsic detector-noise floor. Compared to the individual methods, we observe a larger decrease of the population of retained pulses when applying the combined pre-filter and LAD selection. Without the pre-filter, the LAD is only sensitive to the most pronounced deviations from the expected detector response. After removing the most obvious artifacts with the pre-filter, the LAD can identify more subtle deviations, which results in a more conservative selection of high-confidence pulses.  

\subsubsection{Supervised Extrapolation to Full Datasets}
Once a representative subset of pulses has been labeled, a supervised classifier can be trained to reproduce the high-confidence selection on previously unseen traces without repeating computationally expensive dimensionality reduction and clustering. For the present dataset the random forest classifier (RFC) \cite{breimanRandomForests2001} provided the most favorable trade-off between classification performance, robustness, and computational cost (compared to other classifier architectures like \textit{SGD} \cite{bottouLargeScaleMachineLearning2010} or \textit{Rocket} \cite{dempsterROCKETExceptionallyFast2020}), and was therefore selected. The model is implemented with scikit-learn and trained on raw traces with labels derived by the combined pre-filter and LAD pipeline as targets.

On an evaluation dataset of comparable size and that is independent of the training set, the RFC yields an \textit{AUC} (area under the curve) of $0.931$. We assume that labels found by the LAD are the ground truth for our comparison. At the operating point maximizing the trade-off between precision and recall, the classifier reached a precision of $0.818$, and a recall of $0.934$ (corresponding to a $F_1$ score of $0.872$). The selected RFC population remains close to the combined reference selection in quality (see tab.~\ref{tab:template_quality} and fig.~\ref{fig:LAD comparison}), with a template variation comparable to the LAD-selected population and a reduction of the unexplained variance between the LAD and combined pipeline-selected populations. This demonstrates that the LAD-derived selection can be reliably reproduced on unseen data, and that the RFC can be used to extrapolate the selection to the full dataset.

\subsubsection{Comparison to Template \texorpdfstring{$\chi^2$}{Chi-squared}}
\begin{figure}[ht!]
    \centering
    \includegraphics[width=0.45\textwidth]{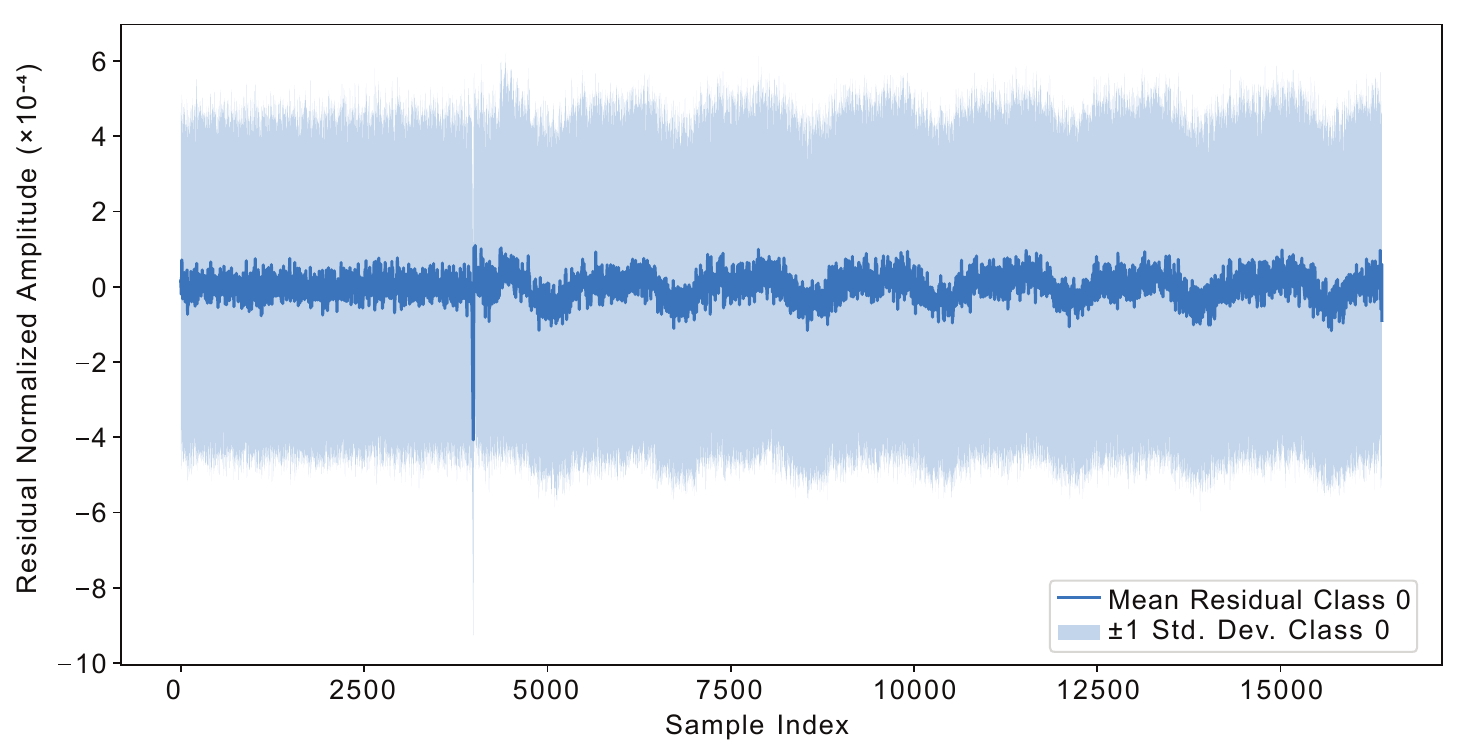}
    \caption{The trace shows the point-wise mean residuals of traces from the LAD-selected population Class~0 (corresponding to traces rejected by the LAD scheme). The residuals are computed after matching the common template to the pulse. The shaded area indicates the point-wise standard deviation of the residuals. One can see the impact of slightly misaligned pulse arrival times (spike around sample index 4000) as well as a structured baseline contamination. This artifact is significantly stronger throughout the entire trace following the trigger as compared to Class~1 (corresponding to the accepted high-confidence pulse population).}
    \label{fig:Class residuals}
\end{figure}
While the LAD is a powerful tool for identifying populations of traces that are statistically incompatible with a representative detector-response template, it rejects a large fraction of traces that are still relevant for downstream feature extraction. In particular, the LAD is intentionally conservative in its selection of a high-confidence template population, which is a necessary step for constructing an optimal filter. However, once a representative template has been established, it is desirable to retain as many traces as possible for the subsequent analysis pipeline. Complementary to the LAD, we could therefore use the template to define a $\chi^2$-like filter that quantifies how alike an individual trace is to the established detector response:
$$
Q_i = \frac{1}{N} \sum_{j=1}^{N}\frac{\left(x_{ij} - \hat{T}_{ij}\right)^2}{\sigma_{n,j}^2}
$$
Here $\sigma_{n,j}$ is the point-wise standard deviation of the noise traces $n_{ij}$, and $x_{ij}$ and $\hat{T}_{ij}$ are once again the $j$-th sample of the $i$-th trace and the corresponding matched template, respectively. We refer to this quantity as a template $\chi^2$ because it is constructed as a noise-normalized sum of squared residuals, analogous to the conventional $\chi^2$ goodness-of-fit statistic. However, it should be emphasized that the template $\chi^2$ is not a formal $\chi^2$ statistic in the strict statistical sense, since the detector noise is temporally correlated. Instead, it is best interpreted as a noise-normalized residual metric that can be used to compare populations and cuts, rather than as an absolute goodness-of-fit probability.

When applied to the training dataset, the acceptance curve of the template $\chi^2$ filter exhibits a pronounced transition followed by a broad retention plateau. We therefore choose the upper edge of the transition, corresponding to \SI{99}{\percent} of the height of the initial transition, as a conservative operating point. At this point, the filter retains a larger fraction of traces (\SI{95.7}{\percent}) compared to the pre-filter (\SI{91.7}{\percent}). Inspecting the traces on which the two methods disagree, we find the template $\chi^2$ filter to retain a population of valid traces (\SI{4.8}{\percent}) that were falsely rejected by the pre-filter~\textendash~mostly on the basis of having a incorrect decay time assigned. Additionally, the template $\chi^2$ filter rejects a small population of traces (\SI{0.7}{\percent}) that were falsely retained by the pre-filter, which are mostly traces that are overlapped by periodic short amplitude spikes. Particularly for experiments with low photon flux, retaining an additional \SI{4.8}{\percent} of valid traces can be significant, showing that the template $\chi^2$ filter can be a useful tool built on top of the results of the LAD and pre-filter methods~\textendash~improving their capability of identifying pulse-like traces for the feature extraction stage.

Finally, the LAD labels provide additional diagnostic value beyond acceptance/rejection. In the presented dataset, they separate two waveform populations with statistically distinguishable characteristics: Class~0 (LAD-rejected), comprising approximately \SI{35}{\percent} of the traces, shows a baseline contamination that is systematically stronger by a factor of approximately $1.7$ compared to traces of Class~1 (LAD-accepted). The residual baseline of Class~0 after subtracting the matched template can be seen in fig.~\ref{fig:Class residuals}. This separation is directly visible in template-compatibility statistics based on explained variance: for Class~0, the mean value is $R^2=0.990$ with a \SI{95}{\percent} confidence interval of $[0.974,\,0.997]$, while for Class~1, the mean is $R^2=0.998$ with a \SI{95}{\percent} confidence interval of $[0.995,\,0.999]$. The distributions are therefore clearly separated, indicating that the two populations are substantially incompatible with a common high-quality template. The origin of this waveform difference is not established within the present analysis and is subject to ongoing investigation; its detailed characterization is beyond the scope of this work.

In contrast, the corresponding noise-weighted template $\chi^2$ distributions show no appreciable separation despite the pronounced structured baseline deviations in the LAD-rejected traces. Specifically, Class~0 yields a mean template $\chi^2$ of $1.029$ (\SI{95}{\percent} CI: $[0.941,\,1.140]$), whereas Class~1 yields $1.027$ (\SI{95}{\percent} CI: $[0.947,\,1.126]$), with strongly overlapping intervals. This result emphasizes the complementary role of both tools: LAD identifies statistically inconsistent waveform populations and contamination classes without requiring an \textit{a~priori} specification of all deviations, while the template $\chi^2$ remains a practical event-level compatibility criterion once a representative template has been established.


\subsection{Example 2: pulse-shape Analysis and Feature Extraction}

The main task of the pulse processing pipeline is to extract relevant features from the raw detector pulses, like the pulse amplitude and exact pulse arrival time relative to the trigger (for simplicity, in the following we refer to it as trigger time). Traditionally, this is achieved using FIR filters like the optimal filter \cite{hengstlerDevelopmentCharacterizationTwodimensional2017} or MWD \cite{herdrichAnwendungKryogenerKalorimeter2023}. The working principle of these methods relies on the exact knowledge of the pulse-shape and noise characteristics. Therefore, they need to be hand-optimized for each pixel and measurement condition, which becomes increasingly impractical for large pixel counts. Additionally, these methods have limited adaptability to changing conditions and complex pulse-shapes that might arise in real-world scenarios.


\subsubsection{Domain Adaptation}

Instead, ML methods, particularly NN, can be utilized to learn the mapping from raw pulses to relevant features directly from the data. This approach allows for automatic adaptation to different pulse-shapes and noise characteristics, making it more robust and scalable. However, as discussed earlier, the lack of labeled data poses a significant challenge for supervised learning methods in this context. To overcome this, we can leverage simulated data to train the model in a supervised manner and subsequently apply transfer learning techniques to adapt the model to real detector data. This so-called \textit{simulation-to-real} domain adaptation approach takes advantage of the fact that the general pulse-shape is well understood from first principles and can be accurately simulated \cite{herdrichAnwendungKryogenerKalorimeter2023}. Using transfer learning methods to bridge the gap between simulation and reality not only allows us to postpone the introduction of real data to a later stage, but also to train on well defined and labeled synthetic data with a wide variety of possible parameters compared to a small set of real sensors (which might only have marginal variations between most elements). 


\subsubsection{MMC Pulse Simulation}

\begin{figure}[ht!]
    \centering
    \includegraphics[width=0.45\textwidth]{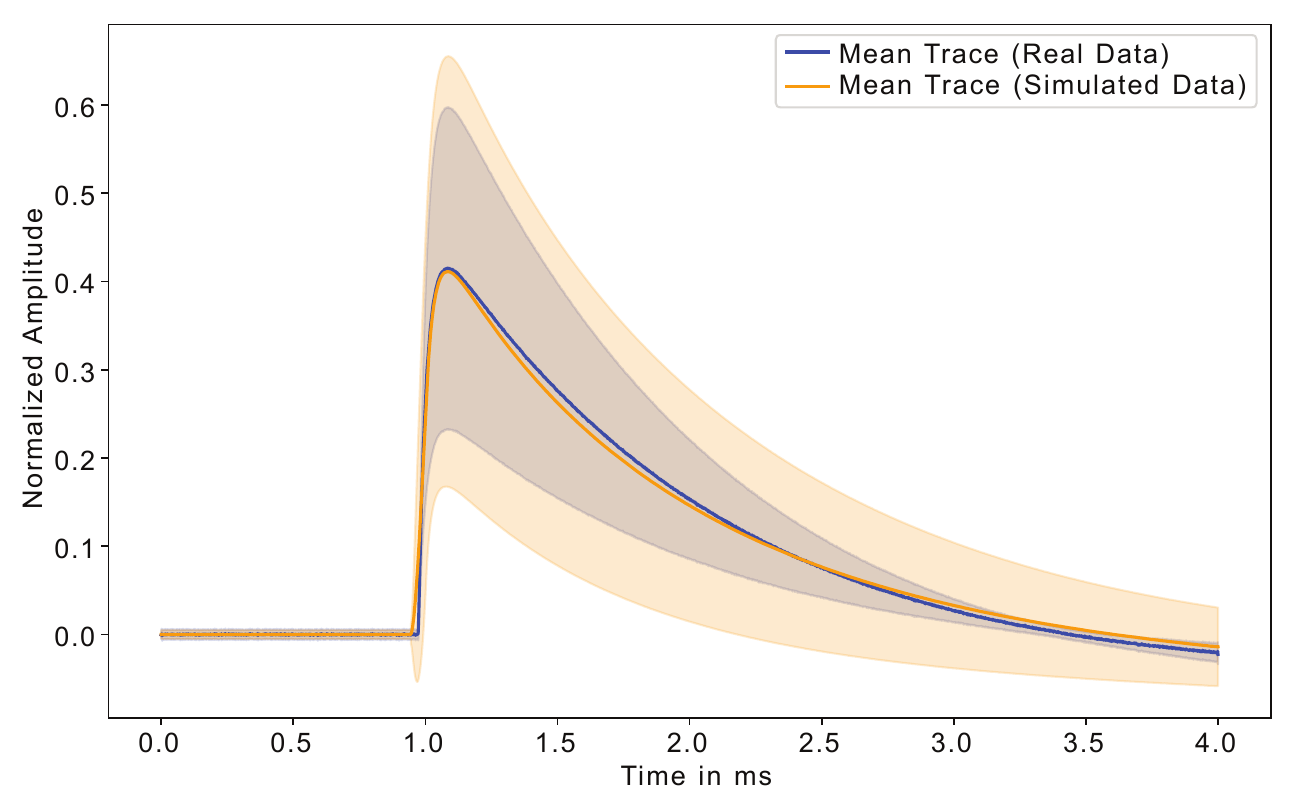}
    \caption{Comparison of the mean pulse-shape of real detector data (blue) and simulated training pulses (orange). The shaded areas indicate the standard deviation of the respective pulse-shapes. The simulation parameters are randomly sampled from predefined ranges to generate a diverse set of pulse-shapes and noise profiles that fully envelop the range of real pulses.}
    \label{fig:Real-Simulation-Pulse Comparison}
\end{figure}

\begin{figure*}[t]
    \centering
    \includegraphics[width=0.87\textwidth]{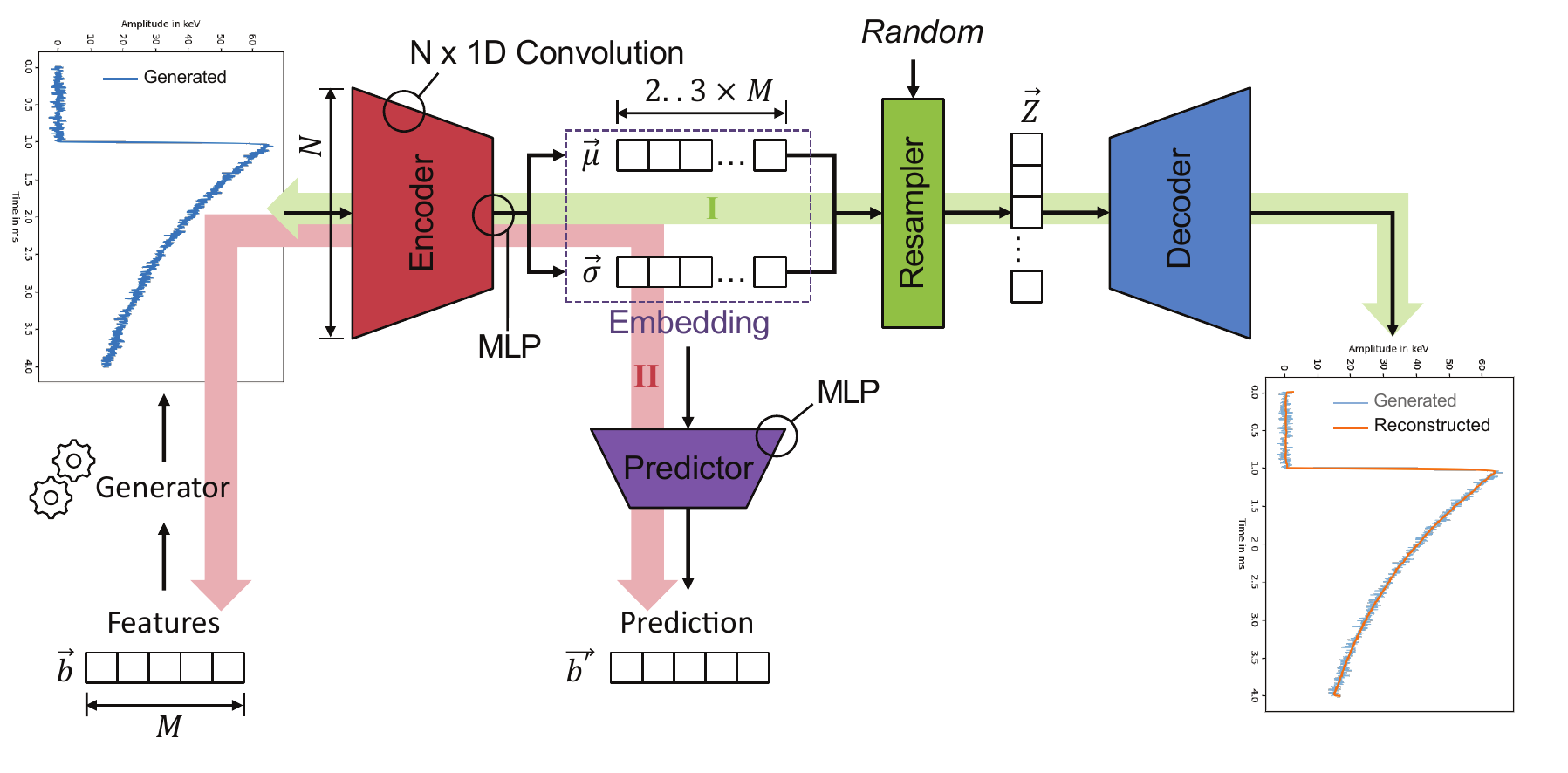}
    \caption{Schematic representation of the CVAE architecture for feature extraction utilizing latent space regression. During the first stage of the training of the network (I, green arrow), sets of randomized pulse features $\vec{b}$ are used to simulate a corresponding MMC pulse (an example is shown in the top left). The encoder stage of the CVAE architecture maps the pulse to a representative distribution in the lower-dimensional latent space (embedding). The distribution $\vec{\mu}, \vec{\sigma}$ is then sampled to generate a latent space vector $\vec{Z}$. It is finally passed to the decoder stage to reconstruct the original pulse (a comparison between the reconstruction and the original pulse is shown in the bottom left). Since during the training we only minimize the difference between the input and output pulses, this step is fully unsupervised. The encoder and decoder learn a representation of the MMC pulse-shape, while all relevant information about the individual pulses are mapped to the embedding. In the second stage of the training (II, red arrow), a dedicated network based on a multi-layer perceptron architecture is trained to map the latent space distribution to the original pulse features $\vec{b}$. This supervised training step allows us to predict the original features of each pulse from its embedding.}
    \label{fig:CVAE MLP Schema}
\end{figure*}

The general pulse-shape of the MMC detector is well understood from first principles and can be described using a simplified model of the detector response and the noise profile:
$$
S(t) = \theta(t-t_0)\cdot A\cdot \left[ -e^{\frac{t-t_0}{\tau_r}} + \left(1 + o\right)\cdot e^{\frac{t-t_0}{\tau_d}} - o \right] + N(t)
$$
Here $S(t)$ is the simulated pulse-shape, $\theta(t-t_0)$ is the Heaviside step function that ensures the pulse starts with delay $t_0$ relative to the trigger time, $A$ is the pulse amplitude, $\tau_r$ and $\tau_d$ are the rise and decay time constants, respectively, and $o$ is an additional parameter, hereafter referred to as the offset parameter, that describes the residual signal offset at the falling edge of the pulse. In the maXs-100 detector data analyzed in this work, the signal falls below the baseline after the main pulse, an effect attributed to the AC coupling between the sensors and the readout electronics. The signal subsequently returns to the baseline on a timescale longer than the sampling window; o therefore parametrizes the undershoot visible at the end of the simulated pulse. To simulate subsample trigger jitter (i.e., the starting point of the pulse does not typically  align with the sampling grid of the ADC), the signal is defined continuous in time and then sampled at discrete points shifted by the corresponding simulated trigger time.

The noise term $N(t)$ is modeled as a combination of white and colored noise:
$$
\text{FFT}\left[N\right](f) = \text{FFT}\left[rnd\right](f) \cdot A_N \cdot \sqrt{A_w^2 + (A_c f^{-\alpha})^2}\cdot e^{-\left(\frac{f}{f_c}\right)^n}
$$
The amplitude of the white noise is given by $A_w$, while the colored noise is characterized by its amplitude $A_c$ and spectral index $\alpha$. The amplitude $A_N$ is used to scale the overall noise amplitude. Finally, $f_c$ is a threshold frequency for a high-frequency roll-off. The steepness of the roll-off is controlled by $n$. The noise term is generated in the frequency domain by transforming normally distributed white noise $rnd$ to the frequency domain and multiplying it with the desired noise spectrum. The resulting noise is then transformed back to the time domain using an inverse Fourier transform.

The simulation parameters are randomly sampled from predefined ranges to generate a diverse set of pulse-shapes and noise profiles. This allows the model to learn a wide variety of possible pulse-shapes and noise characteristics, improving its generalization capabilities when applied to real detector data (see fig.~\ref{fig:Real-Simulation-Pulse Comparison}). The parameters are set up to be comparable to the characteristics of the pixels found in real maXs-100 detectors and set to \SIrange{20}{30}{\percent} fluctuations around the mean values for the training data set. This is about one order of magnitude larger than the typical per-pixel fluctuation of these parameters in the real detector.

\begin{figure*}[t]
    \centering
    \includegraphics[width=0.90\textwidth]{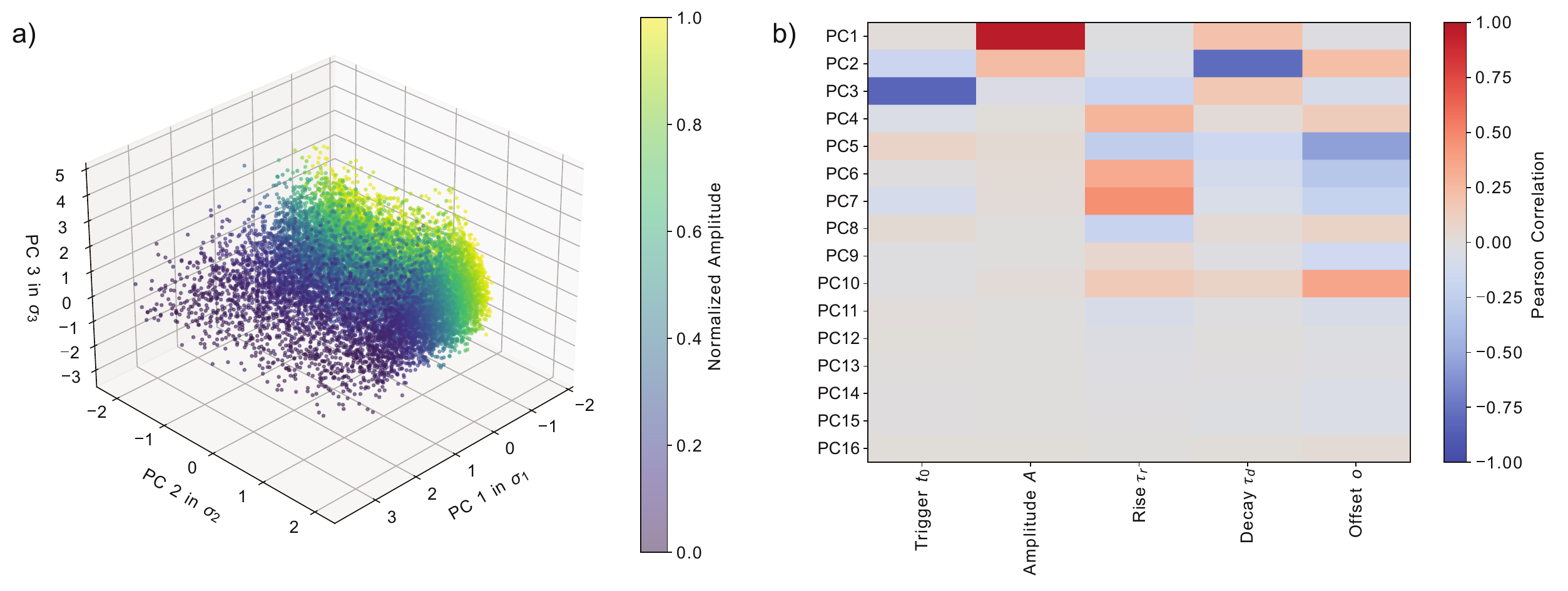}
    \caption{\textit{a)} The left plot shows the embedding of the $16$-dimensional latent space representation of simulated pulse reduced via PCA to $3$ dimensions. Each point represents a simulated pulse. The color indicates the simulated amplitude of each pulse. The axes are the three principal-component coordinates and are given in units of their standard deviation. \textit{b)} The right panel shows the Pearson correlation matrix between the simulation parameters and all $16$ principal components. The correlation values are given in the range of $[-1, 1]$, with $1$ indicating a perfect positive correlation and $-1$ indicating a perfect negative correlation. The first three principal components show the strongest correlation with the simulation parameters, particularly with the pulse amplitude, trigger time, and decay time. The strong correlation, for example, between the amplitude and the first principal component explains the alignment of the color gradient in \textit{a)} with the first principal component.}
    \label{fig:Latent Space PCA}
\end{figure*}

\subsubsection{Unsupervised Reconstruction and Latent Space Regression}
The first step in the feature extraction process is to learn a compact representation of the pulse-shapes via unsupervised reconstruction - effectively also acting as a dimensional reduction step \cite{hintonReducingDimensionalityData2006}. This is achieved using a \textit{CVAE} (Convolutional Variational Auto-Encoder)-based architecture \cite{kingmaAutoEncodingVariationalBayes2022} (see fig.~\ref{fig:CVAE MLP Schema}). The CVAE consists of an encoder and a decoder network that work together to compress the input pulse into a lower-dimensional latent space representation and then reconstruct it back to the original shape. The encoder consists of a hierarchy of one-dimensional residual convolutional blocks that progressively downsample the input pulse while increasing the number of feature channels. The extracted feature representation is flattened and mapped to the mean and variance of a latent Gaussian distribution. For the reconstruction, a latent space vector is randomly sampled from this distribution and fed into the decoder. The objective of the CVAE is to minimize the difference between the input and reconstructed pulses, effectively learning the structure of the input data and finding a representation of all the relevant information about each individual pulse in the latent space embedding. Compared to a regular auto-encoder, the CVAE learns parameter distributions in the latent space and samples randomized output values, allowing for better generalization and robustness to noise. We allow more dimensions in the latent space than strictly necessary to give the network some freedom to expand relevant features. This also helps to prevent over-compression and loss of important information.

Analyzing the latent space provides direct insight into which pulse properties the network has encoded. To make this structure visible, we perform a PCA of the latent representations (see fig.~\ref{fig:Latent Space PCA}): The left side of the figure shows the PCA ($n=3$) of a set of pulses embedded into the latent space. Each point corresponds to a simulated pulse, colored by its corresponding known simulation amplitude. The color gradient shows a clear correlation between the amplitude parameter and the first principal component. When we calculate the correlation between the simulation parameters and all $16$ principal components of the latent space (see fig.~\ref{fig:Latent Space PCA} b) we can study how the parameters are represented in the latent space. The dominant principal components align with the simulation parameters that contribute the largest variance to the pulse morphology, most prominently pulse amplitude as well as trigger timing and decay time variations. This result is particularly informative because the CVAE was not trained with explicit parameter labels. Instead, the correspondence emerges purely from the reconstruction objective, i.e., from learning an embedding that preserves the relevant structure of the input traces. The observed alignment therefore indicates that the model has autonomously organized the latent space along physically meaningful degrees of freedom.

In a second stage, a simple neural network based on a \textit{MLP} (Multi-Layer Perceptron) architecture can then be trained on the simulated data to learn the mapping from the latent space representation to the relevant features. This model is trained using the known labels from the simulation and can then be applied to real data later, if the simulation parameters are chosen to cover the expected real-world variations well enough. The parameter prediction network consists of an input projection followed by three fully connected residual blocks. Separate regression heads are used for each physical parameter, allowing each target to learn a dedicated mapping from the shared latent representation. The technique of feature reconstruction from latent space embedding is known as latent space regression \cite{weisserUnsupervisedParameterEstimation2023} and has been successfully applied in various contexts.

Both architectures have been implemented using the PyTorch library. The CVAE is trained using a combination of reconstruction loss and a regularization term that encourages the latent space to follow a predefined distribution, typically a multivariate Gaussian. The MLP is trained using mean squared error loss between the predicted and true feature values. Hyper-parameters for both models are optimized via cross-validation to maximize performance.

\subsubsection{Comparison between FIR filter and neural network}
To compare both approaches under controlled but realistic conditions, we evaluate amplitude reconstruction on simulated datasets spanning increasing levels of pulse-shape and noise variability. The comparison is performed by calculating the residuals between the predicted and true amplitude for each pulse. For the ML pipeline, a small systematic amplitude bias in the residuals is removed using a \textit{LOWESS} (LOcally WEighted regreSSion \cite{Cleveland1979}) calibration, after which the residual distribution is approximately Gaussian over the relevant dynamic range. As a baseline we use the MWD implementation, because it is the established and well-characterized method in the maXs analysis chain for non-ideal operating conditions \cite{herdrichAnwendungKryogenerKalorimeter2023}; although the optimal filter can approach the theoretical limit for stable pulse-shapes, its assumptions are less representative for the strongly varying scenarios considered here. In a nearly ideal single-channel setting without parameter variation, the FIR filter reaches \SI{72}{\electronvolt} FWHM, close to the expected optimal-filter limit of \SI{69}{\electronvolt}, whereas the ML estimator yields \SI{133}{\electronvolt}. However, this regime does not reflect realistic detector operation.
\begin{figure}[ht!]
    \centering
    \includegraphics[width=0.45\textwidth]{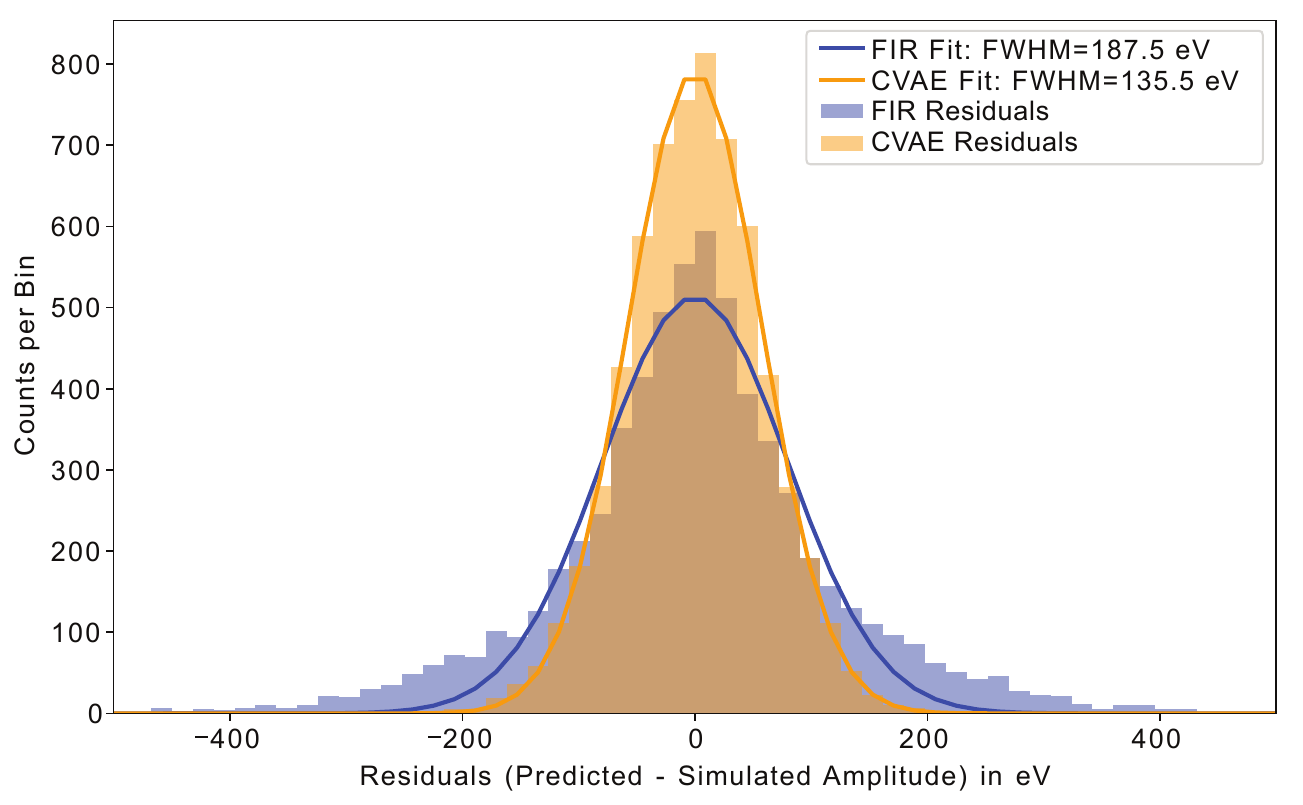}
    \caption{A set of simulated data was analyzed using both the traditional MWD filter (FIR, blue bars) and the ML-based feature extraction method (CVAE, orange bars). For both methods the distribution of residuals computed between the predicted and known amplitude used in the simulation is shown. The FWHM values given in the legend refer to Gaussian fits to the residual distributions and the blue and orange curves show the matched normal distributions respectively. The ML-based method shows a narrower distribution, indicating improved accuracy in amplitude estimation compared to the traditional FIR filter. Due to a slight amplitude sensitivity of the FIR filter, the residuals show a non-Gaussian broadening towards larger deviations.}
    \label{fig:CVAE FIR Comparison}
\end{figure}

When channel-to-channel parameter fluctuations are introduced at levels representative of measured per-pixel distributions, the FIR resolution degrades strongly while the ML result remains comparatively stable: we obtain \SI{136}{\electronvolt} (ML) versus \SI{188}{\electronvolt} (FIR) FWHM (see fig.~\ref{fig:CVAE FIR Comparison}). The relative separation between both methods is therefore larger in this realistic-variation case. For a deliberately selected pixel with large noise contributions, both methods broaden substantially but ML still retains an advantage (\SI{563}{\electronvolt} versus \SI{627}{\electronvolt} FWHM, with an estimated optimum of \SI{299}{\electronvolt} without pulse-shape variations). In an additional stress test, applying the full training-variation range to the simulation increases the ML width by only a factor of about $\num{1.5}$, whereas the FIR-based methods no longer provide a usable reconstruction. In all tests the FIR residuals also show excess tail broadening relative to a Gaussian model, consistent with energy-dependent sensitivity to pulse-shape changes, whereas ML residuals remain closer to Gaussian. For all tested parameter distributions the same network was used without changes, while the FIR filter had to be tuned to the specific characteristics. Only when we de-tune a parameter that drastically changes the noise profile outside of the learned training set~\textendash~for example, changing the color of the noise spectrum from $\alpha=\num{-0.7}$ to $\num{-0.3}$~\textendash~the performance of the ML filter degrades more significantly than that of the comparable FIR filter.

\subsubsection{Transfer to Real-World Data}
After training on simulated data, the models can be applied to real detector pulses. The real-world benchmark is based on samples recorded using a maXs-100 family detector in our recent beam time campaign in 2025 at the CRYRING@ESR of the GSI/FAIR campus. The pulses were sampled with $2^{14}$ total samples per trace, which had to be downsampled to $2^{12}$ samples using a polyphase-based filtering to match the input size of the ML filter. We have chosen an ADC channel that yielded high counting statistics and limited the data to a calibration spectrum recorded in the span of one day. During the calibration an Americium-241 and a Gadolinium-153 radiation source were placed in front of the detector, with a thin film of tin and an air-gap in between to attenuate lower energy atomic radiation. Photons emitted by the radioactive sources cover energies in the entire spectral range of the detector (around \SIrange{10}{100}{\kilo\electronvolt}). In total, the dataset contained around $16000$ samples, which was reduced to $14000$ after applying the pre-filter pipeline for valid pulse selection. The same data-selection and calibration protocols were applied to the FIR-based reference pipeline and the ML-based analysis to ensure a consistent comparison.
\begin{figure}[ht!]
    \centering
    \includegraphics[width=0.45\textwidth]{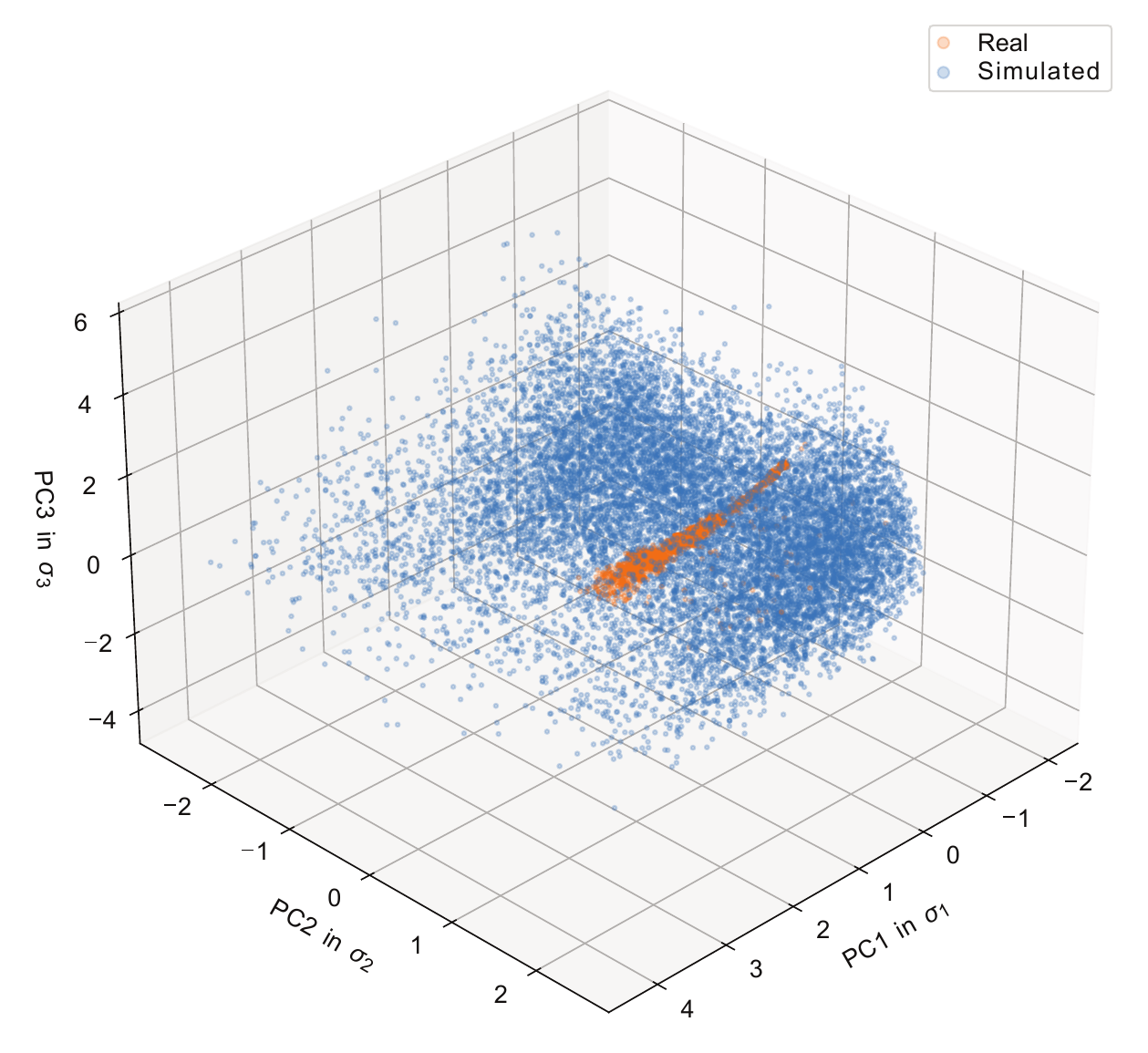}
    \caption{Comparison of the $16$-dimensional latent space embedding of simulated and real pulses reduced via PCA to $3$ dimensions. Each point represents a pulse taken from the real-world dataset (orange) or the simulation (blue). The axes are the principal-component coordinates given in units of their standard deviation (across the combined set of samples). By design of the training setup, the real pulses occupy a much smaller region fully embedded within the simulated pulse distribution. This indicates that the set of synthetic pulses covers all relevant variations of the real detector pulses, making the transfer learning approach feasible.}
    \label{fig:CVAE Real Data Performance}
\end{figure}

\paragraph*{Network Fine-Tuning}
Due to differences between the simulated and real data distributions, a direct application might lead to suboptimal performance. To address this, a fine-tuning step of the NN model is performed using a subset of real data (around \SI{40}{\percent}). This involves freezing the weights of the MLP and only updating the weights of the CVAE during training. This allows the encoder to adapt to the specific characteristics of the real data while preserving the learned mapping from latent space to features. The fine-tuning is performed using a reduced learning rate and a limited number of epochs to prevent overfitting. The real detector data is pre-filtered by the classifier to avoid training on artifacts.

\paragraph*{Latent Space Embedding and Reconstruction Error}
When comparing the embedding of simulated (as used for the network training) and real pulses in the latent space (see fig.~\ref{fig:CVAE Real Data Performance}) it becomes apparent that the real pulses occupy a much smaller region in the latent space that is completely contained within the simulated pulse distribution. This indicates that the simulation is able to cover all variations of the real detector pulses relevant to the present analysis, making the transfer learning approach feasible in principle. Before fine-tuning, the reconstruction error (MSE) of the real-world data is a factor of $1.6$ higher than for the simulated pulses with similar pulse parameter distributions. After applying the fine-tuning step the reconstruction error drops to the same level.

\paragraph*{Comparison of Spectral Resolution}
\begin{figure}[ht!]
    \centering
    \includegraphics[width=0.45\textwidth]{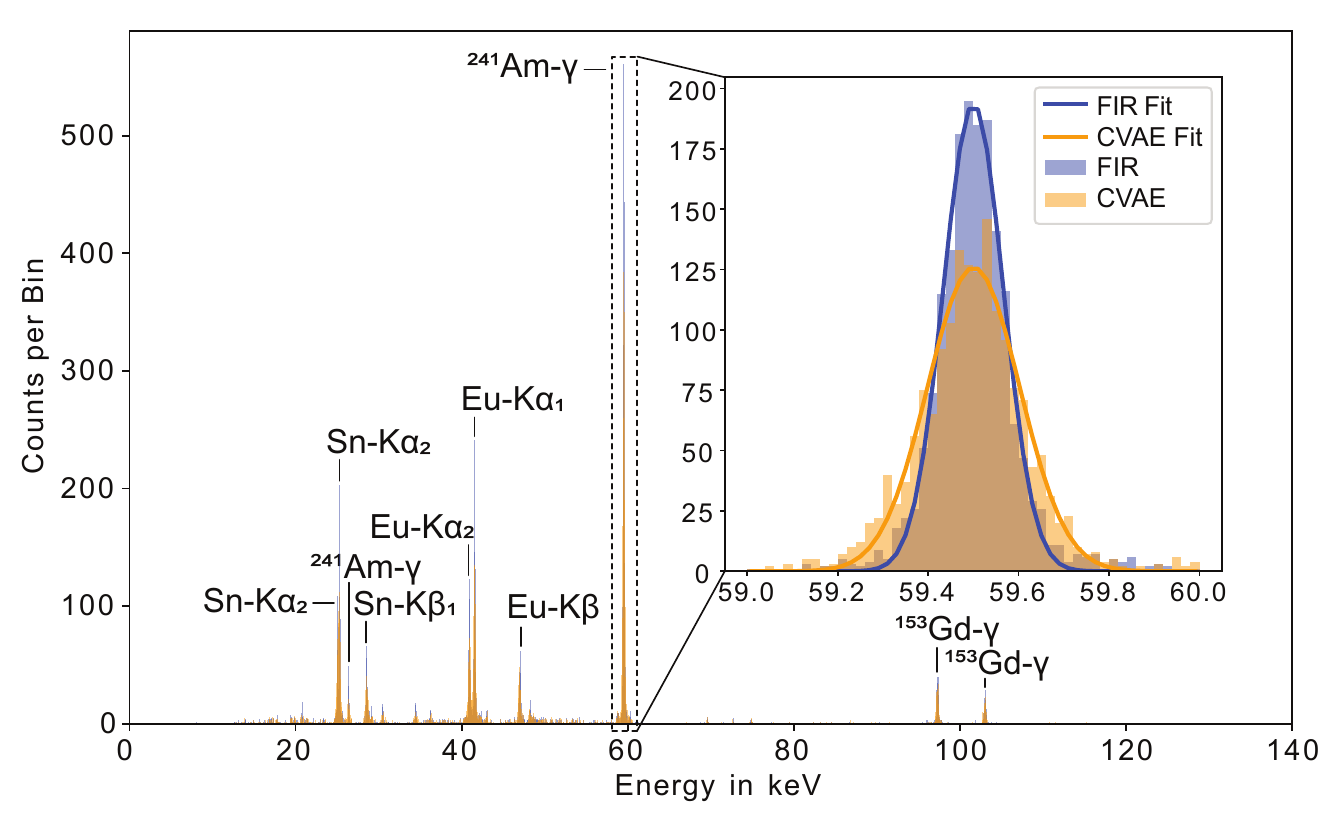}
    \caption{Comparison of energy spectra obtained from ML-based feature extraction (CVAE, orange) and traditional MWD filtering (FIR, blue) for a single detector channel using Am-241 and Gd-153 calibration sources and a tin film for blocking lower energy atomic radiation. $\gamma$-lines from the sources as well as characteristic X-rays from their daughter nuclides are visible, as well as fluorescence from the Sn shield. The spectra show good agreement in peak positions and linearity across the entire spectrum. The data was temperature corrected and calibrated using the \SI{59.5}{\kilo\electronvolt} $\gamma$-line of Americium. The inset shows a zoom on the \SI{59.5}{\kilo\electronvolt} peak. The FWHM of the peak is calculated by matching a Gaussian distribution to the data which yields \SI{161}{\electronvolt} in the case of the FIR filter (blue curve) and \SI{240}{\electronvolt} for the ML-based filter results (orange curve).}
    \label{fig:Energy Spectra Comparison}
\end{figure}
Before applying the gain-drift correction to the data, the FIR-based MWD filter achieves an FWHM of \SI{1133}{\electronvolt} for the \SI{59.5}{\kilo\electronvolt} $\gamma$-line of Americium, while the ML-based method achieves an FWHM of \SI{1198}{\electronvolt} without fine-tuning of the network applied. After applying the gain-drift correction, the FIR filter result improves to \SI{161}{\electronvolt} FWHM. This is compatible with the simulation result of \SI{188}{\electronvolt} FWHM (most likely the simulation overestimates the parameter variation compared to the real pulses). However, the ML result only improves to \SI{276}{\electronvolt} FWHM. After applying the fine-tuning step on a small subset of the real data, the ML result improves to \SI{240}{\electronvolt} FWHM (see fig.~\ref{fig:Energy Spectra Comparison}). In the non-temperature corrected case, the ML result improves to \SI{1100}{\electronvolt} FWHM, which slightly outperforms the FIR filter. The following discussion addresses the comparatively limited improvement of the ML approach relative to the simulation results, as well as the substantial difference between the ML and FIR performance after gain-drift correction.

\paragraph*{Simulation-to-Real Domain Mismatch}
To explain the discrepancy between the simulation and real data performance, we analyze the latent space embedding of the real pulses. We prepared a dataset by individually fitting the real pulses to the MMC model and simulating pulses with identical parameters. After embedding both the real and simulated pulses into the latent space, we find that the real pulses are shifted in some latent space dimensions compared to the simulated ones. This shift could not be explained through deliberate de-tuning any of the simulation parameters~\textendash~which indicates that the simulation model is not able to perfectly capture the real detector response. Differences between the real pulses to the model degrade the ML performance because they don not get mapped to the same location in the latent space as the simulated pulses, which prevents the MLP from accurately predicting the parameters. A potential candidate for this effect is, e.g., the INL (Integral Non-Linearity) of the readout ADCs which systematically distorts the shape of the pulses. This is further supported by the fact that increasing the training depth of the ML worsens the performance on real data, which indicates that the network over-fits to the simulation model and thus loses its ability to generalize to real data. The fine-tuning step partially mitigates this effect by allowing the ML to adapt to the real data distribution.

\paragraph*{Temperature-Dependent Pulse Variations and Gain-Drift Correction}
Furthermore, the energy resolution of the ML filter result seems to be on the same level as the FIR result (about \SI{3}{\percent} better) before applying the gain-drift correction. After the application of the linear gain-drift correction it is outperformed by the FIR filter. This indicates that the gain-drift correction is less effective on the ML filtered data. To explain this, we analyzed the correlation between all pulse parameters and the chip temperature in the present dataset. We found a linear dependence of the pulse's rise time and offset parameter on the temperature. This means that not only the amplitude but also the pulse-shape varies with temperature. The MWD is static and translates the change into a systematic shift of the amplitude. This shift is mostly corrected out by the linear gain-drift correction itself. The neural network however is trained on a wide variety of pulse-shapes and can accurately map the amplitude more independently of the pulse-shape. As a result, the ML filter is already partially mitigating the temperature-dependent amplitude shift implicitly. This is verified by the fact that the $R^2$-value of the linear correlation between amplitude and temperature is reduced by \SI{18}{\percent} between the FIR and ML results (relative to the FIR). Thus, by partially removing the temperature-dependent amplitude variation, the ML filter reduces the amount of temperature dependence remaining for the subsequent linear gain-drift correction to remove. To further test this hypothesis, we investigated the sensitivity of the gain-drift correction to the tails of the calibration-pulse amplitude distribution. Prior to applying the temperature correction, a symmetric fraction of the most extreme calibration events was successively excluded from the distribution, after which the gain-drift correction and energy-resolution analysis were repeated. Assuming that amplitude variations within the same $\gamma$-peak are predominantly caused by detector temperature drifts, the distribution tails isolated by this procedure represent their temperature-dependent contribution. The ML-filtered data showed a significantly stronger improvement in energy resolution upon removing these events~\textendash~approximately \SI{1.9}{eV} per removed percentile in the tail of the amplitude distribution, compared to \SI{1.2}{eV} for the FIR filter. This difference demonstrates that the two filters respond differently to the temperature-dependent amplitude variations. In combination with the reduced amplitude-temperature correlation observed for the ML filter, the stronger sensitivity to the removal of the residual tails supports the hypothesis that the ML filter already partially compensates the temperature-dependent amplitude variation, thereby reducing the amount of variation that can subsequently be removed by the linear gain-drift correction. This effect could be mitigated in future work by explicitly studying the correlation between the latent space embedding and the temperature to include the correction directly into the ML architecture itself. This would potentially allow us to remove the temperature correction step from the analysis pipeline and thus reduce the number of processing steps and potential sources of error. A dedicated measurement campaign is scheduled in the future to produce a suitable dataset.

\paragraph*{Current Capabilities and Remaining Challenges}
These results do not yet establish a general replacement for the traditional FIR-based analysis chain. Rather, they demonstrate that a simulation-trained model can recover physically relevant pulse features and remain competitive under realistic detector variability, while further improvement of the simulation model and domain adaptation is required to reach the full performance of the optimized conventional analysis. To mitigate the observed shift between the simulated and real data distributions in the latent space, several strategies can be explored. One pathway is improving the simulation model to better match the real detector response. This could, for example, be accomplished by using a \textit{GAN} (Generative Adversarial Network) \cite{madryDeepLearningModels2019} to generate simulated data that is closer to real-world data. Furthermore, more involved domain-adaptation methods, such as domain-adversarial training \cite{JMLR:v17:15-239} or domain-separation networks with private and shared representations \cite{NIPS2016_45fbc6d3}, could be explored to better align the simulated and real data distributions in the latent space. Finally, a more robust regression model that could handle small shifts in the latent space representation could be investigated to improve the performance of the feature extraction on real data.

At least before the gain-drift correction, the ML-based method shows comparable energy resolution to the FIR filter. This result is particularly remarkable, considering that the ML models were trained solely on simulated data and only fine-tuned on a small subset of real data, while the FIR filter had to be hand-optimized to match the characteristics of the selected detector channel. Whereas this optimization has to be repeated for each detector pixel and measurement, the ML approach could potentially be extended to an arbitrary number of sensors in an automated manner, provided that their individual characteristics are sufficiently represented by the training data. The results should therefore be interpreted as a proof of principle for simulation-to-real domain adaptation in MMC pulse analysis, not as a demonstration of a fully mature replacement for the conventional analysis chain. However, this result is already very promising for the implementation of online processing schemes for real-time monitoring and feedback during the experiment, where both the computational efficiency and flexibility of ML methods can be highly advantageous and are more important than the absolute performance of the method. 


\section{Conclusion and Outlook}

When applying ML methods to a new field, like the processing of MMC detector signals, several important considerations must be taken into account to ensure successful integration and reliable results. In general, care has to be taken to find applications where ML actually makes sense from the standpoint of task-specific justification. Therefore, we have provided two example applications in the pulse processing pipeline where ML methods could offer advantages as complementary tools to the existing pipeline. The chosen architectures are interpretable in terms of latent space structure, computationally efficient, well-established and theoretically grounded. Verification is achieved through cross-validation on simulated data, consistency checks with known physical behavior, and performance benchmarking against traditional methods.

The results presented in this work show that ML can provide useful and scalable support for MMC pulse processing in two complementary settings. First, the label auto-discovery pipeline based on PCA, UMAP, and HDBSCAN provides a data-driven and largely unsupervised approach to pulse classification and artifact rejection. It identifies waveform populations that are statistically inconsistent with the nominal detector response and thereby yields cleaner template populations without requiring manual labeling. This is particularly valuable in the context of large and highly pixelated detector arrays, where conventional handcrafted selection rules become increasingly difficult to maintain.

Second, the CVAE-based latent-space regression approach demonstrates that physically relevant pulse features can be recovered from raw detector traces using simulation-trained models and limited transfer to real data. In simulations, the method remains competitive across realistic pulse-shape and noise variations and is less reliant on per-channel tuning than the conventional FIR approach. On measured data, the approach remains promising and reaches performance comparable to the optimized FIR baseline before final gain-drift correction; however, the residual mismatch between the simulated and measured pulse distributions still limits full parity with the conventional chain including all corrections.

Taken together, these results suggest that ML should be viewed as a complementary and scalable component of the MMC analysis pipeline rather than as a universal replacement for the established methods. The main benefit lies in automation, robustness to parameter variations, and reduced need for per-channel manual optimization, while the conventional FIR-based pipeline remains the benchmark for the final energy-resolution performance in the present implementation.

Future work will focus on several aspects to further improve and expand the presented methods. The enhancement of the feature extraction stage could be achieved by further upgrading the simulation data source. Incorporating more real-world data into the training process can help the models better capture the complexities of actual detector signals. Further network improvements based on domain adaptation theory could also help bridge the gap to traditional filters.

Furthermore, exploring how to improve gain-drift correction utilizing ML tools could be beneficial. Therefore, time series analysis via \textit{LSTM} (Long Short-Term Memory) \cite{hochreiterLongShortTermMemory1997} networks could be investigated to model and correct for temporal drifts and discontinuities in the detector response. Finally, the hardware optimization of the read-out and amplification SQUIDs could be targeted by a bespoke Bayesian optimization framework (see, for example, J. Duris et al.\cite{durisBayesianOptimizationFreeElectron2020}). This would allow for automatic tuning of the hardware parameters to achieve optimal performance based on real-time feedback from the detector signals. 


\section{Author Declarations}


\subsection{Conflict of Interest}

The authors have no conflicts to disclose.


\subsection{Author Contributions}

M.O. Herdrich: Conceptualization, Methodology, Software, Validation, Formal analysis, Investigation, Data curation, Visualization, Writing - original draft, Writing - review and editing.

T. Mattis: Conceptualization, Methodology, Validation, Writing - review and editing.

G. Weber: Support and data provision for the evaluation of the presented methods, Methodology, Writing - review.

D.A. Schnauß-Müller: Support and data provision for the evaluation of the presented methods, Writing - review.

J.H. Walch: Support and data provision for the evaluation of the presented methods, Writing - review.

Th. Stöhlker: Conceptualization, Supervision, Funding acquisition, Project administration, Writing - review and editing.


\subsection{Acknowledgments}

The authors gratefully acknowledge the colleagues at the HI-Jena, the GSI, and the group of Prof. Dr. C. Enss at the Kirchhoff Institute for Physics, Heidelberg, Germany for providing experimental data used for demonstration and evaluation purposes in this study. The authors further thank the colleagues of the Kirchhoff Institute for Physics for their collaboration in developing the MMC detector system as well as their support with operating the detectors during the experiments. The publication is funded by the Open Access Publishing Fund of GSI Helmholtzzentrum für Schwerionenforschung.


\subsection{Availability of data}

The data that support the findings of this study are available from the corresponding author upon reasonable request.


\nocite{*}
\bibliography{aipsamp}

\end{document}